\documentclass[
 reprint,twocolumn,
 amsmath,amssymb,
 aps,
]{revtex4-1}

\usepackage{epsfig,bm,epsf,graphics}
\usepackage{graphicx}
\usepackage{dcolumn}
\usepackage{bm}

\begin{document}
\preprint{APS/123-QED}

\title{Theory of magnons in spin systems with Dzyaloshinskii-Moriya interaction}

\author{Sahbi El Hog\footnote{sahbi.el-hog@u-cergy.fr} and H. T. Diep\footnote{diep@u-cergy.fr, corresponding author}}
\affiliation{%
Laboratoire de Physique Th\'eorique et Mod\'elisation,
Universit\'e de Cergy-Pontoise, CNRS, UMR 8089\\
2, Avenue Adolphe Chauvin, 95302 Cergy-Pontoise Cedex, France.\\
 }%
\author{Henryk Puszkarski\footnote{henpusz@amu.edu.pl}}
\affiliation{Surface Physics Division, Faculty of Physics, Adam Mickiewicz University,\\
ul. Umultowska 85, 61-614 Poznan, Poland.}

\date{\today}

\begin{abstract}
We study in this paper magnetic properties of a system of quantum Heisenberg spins interacting with each other via a ferromagnetic exchange interaction $J$ and an in-plane Dzyaloshinskii-Moriya interaction $D$.  The non-collinear ground state due to the competition between $J$ and $D$ is determined. We employ a self-consistent Green'function theory
to calculate the spin-wave spectrum and the layer magnetizations at finite $T$ in two and three dimensions as well as in a thin film with surface effects.  Analytical details and the validity of the method are shown and discussed.
\vspace{0.5cm}
\begin{description}
\item[PACS numbers: 75.25.-j ; 75.30.Ds ; 75.70.-i ]
\end{description}
\end{abstract}

\pacs{Valid PACS appear here}
\maketitle


\section{Introduction}

The Dzyaloshinskii-Moriya (DM) interaction was proposed to explain the weak ferromagnetism which was observed in antiferromagnetic Mn compounds.   The phenomenological Landau-Ginzburg model introduced by I. Dzyaloshinskii \cite{Dzyaloshinskii} was microscopically derived by T. Moriya \cite{Moriya}. The interaction between two spins $\mathbf S_i$ and $\mathbf S_j$ is written as

\begin{equation}\label{eq1}
 \mathbf D_{i,j}\cdot \mathbf S_i\wedge\mathbf S_j
\end{equation}
where $\mathbf D_{i,j}$ is a vector which results from the displacement of non magnetic ions located between $\mathbf S_i$ and $\mathbf S_j$,  for example in Mn-O-Mn bonds. The direction of  $\mathbf D_{i,j}$ depends on the symmetry of the displacement \cite{Moriya}.  The DM interaction is antisymmetric with respect to the inversion symmetry.

There has been a large number of investigations on the effect of the DM interaction in various materials, both experimentally and theoretically for weak ferromagnetism in perovskite compounds (see references cited in Refs. \onlinecite{Sergienko,Ederer}, for example). However, the interest in the DM interaction goes beyond the weak ferromagnetism: for example, it has been recently shown in various works that the DM interaction is at the origin of topological skyrmions \cite{Maleyev,Lin,Bogdanov,Rossler,Muhlbauer,Yu1,Yu2,Seki,Adams,Heurich,Wessely,Jonietz} and new kinds of magnetic domain walls \cite{Heide,Rohart}.  The increasing interest in skyrmions results from the fact that skyrmions may play an important role in the electronic transport which is at the heart of technological application devices \cite{Fert2013}.

In this paper, we are interested in the spin-wave (SW) properties of a system of spins interacting with each other via a DM interaction in addition to the symmetric isotropic Heisenberg exchange interaction.  The competition between these interactions gives rise to a non-collinear spin configuration in the ground state (GS). Unlike helimagnets where the helical GS spin configuration results from the competition between the symmetric nearest-neighbor (NN) and next-nearest neighbor (NNN)  interactions \cite{Yoshimori,Villain59}, the DM interaction, as said above, is antisymmetric. This gives rise to a non trivial SW behavior as will be seen below. Note that there has been a number of works dealing with the SW properties in DM systems \cite{Puszkarski,Zakeri,Wang,Stashkevich,Moon}.

This paper is organized as follows. Section \ref{GS} is devoted to the description of
the model and the determination of the GS. Section \ref{GF} shows the
formulation of our self-consistent Green's function (GF) method. Section \ref{2D3D} shows results on the SW spectrum and the magnetization in two dimensions (2D) and three dimensions (3D).
The case of thin films with free surfaces is shown in section \ref{results} where layer magnetizations at finite temperature ($T$) and the thickness effect are presented.
Concluding remarks are given in section \ref{concl}.

\section{Model and ground state}\label{GS}

We consider a thin film of simple cubic (SC) lattice of $N$ layers stacked in
the $y$ direction perpendicular to the film surface.  For the reason which is shown below, we choose
the film surface as a $xz$ plane.
The Hamiltonian is given by

\begin{eqnarray}
\mathcal H&=&\mathcal H_e+\mathcal H_{DM}\label{eqn:hamil1}\\
\mathcal H_e&=&-\sum_{\left<i,j\right>}J_{i,j}\mathbf S_i\cdot\mathbf S_j\label{eqn:hamil2}\\
\mathcal H_{DM}&=&\sum_{\left<i,j\right>}\mathbf D_{i,j}\cdot \mathbf S_i\wedge\mathbf S_j
 \label{eqn:hamil3}
\end{eqnarray}
where $J_{i,j}$ and $\mathbf D_{i,j}$ are the exchange and DM interactions, respectively,
between two Heisenberg spins $\mathbf S_i$ and $\mathbf S_j$ of magnitude $S=1/2$
occupying the lattice sites $i$ and $j$.

For simplicity, let us consider the case where the in-plane and inter-plane exchange interactions between NN
are both ferromagnetic and denoted by $J_{\shortparallel}$ and $J_\bot$, respectively.
The DM interaction is supposed to be between NN in the plane with a constant $D$.
Due to the competition between the exchange $J$ term which favors the collinear configuration, and the
DM term which favors the perpendicular one, we expect that the spin $\mathbf S_i$ makes
an angle $\theta_{i,j}$ with its neighbor $\mathbf S_j$.
Therefore, the quantization axis of $\mathbf S_i$ is not the same as that of $\mathbf S_j$.
Let us call
$\hat\zeta_{i}$ the quantization axis of $\mathbf S_i$ and $\hat\xi_i$ its perpendicular axis in the $xz$ plane.
The third axis $\hat \eta_i$, perpendicular to the film surface, is chosen in such a way to make
($\hat\xi_i,\hat\eta_i,\hat \zeta_i$) an orthogonal direct frame.
Writing $\mathbf S_i$ and $\mathbf S_j$
in their respective local coordinates, one has
\begin{eqnarray}
\mathbf S_{i}&=&S_i^x\hat \xi_i + S_i^y\hat \eta_i+ S_i^z\hat \zeta_i\label{local1}\\
\mathbf S_{j}&=&S_j^x\hat \xi_j + S_j^y\hat \eta_j+ S_j^z\hat \zeta_j\label{local2}
\end{eqnarray}
We choose the vector $\mathbf D_{i,j}$ perpendicular to the $xz$ plane, namely
\begin{equation}\label{Ddef}
\mathbf D_{i,j}=D e_{i,j}\hat\eta_i
\end{equation}
where $e_{i,j}$ =+1 (-1) if $j>i$ ($j<i)$ for NN on the $\hat \xi_i$ or $\hat \zeta_i$ axis.
Note that $e_{j,i}=-e_{i,j}$.

\begin{figure}[ht!]
\centering
\includegraphics[width=8cm,angle=0]{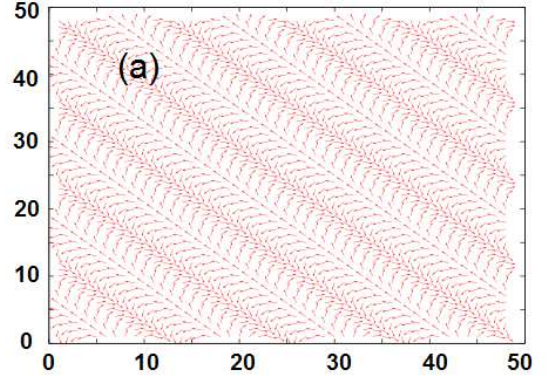}
\includegraphics[width=8cm,angle=0]{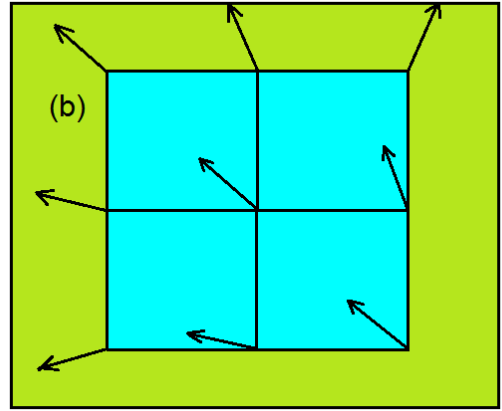}
\caption{(a) The ground state is a planar configuration on the $xz$ plane. The figure shows the case where
$\theta=\pi/6$ ($D=-0.577)$, $J_{\shortparallel}=J_\bot =1$ using
the steepest descent method ; (b) a zoom is shown around a spin with its nearest neighbors. \label{ffig1}}
\end{figure}

To determine the GS, the easiest way is to use the steepest descent method:  we calculate  the local field
acting on each spin from its neighbors and we align the spin in its local-field direction to minimize its energy.
Repeating this for all spins and iterating many times until the convergence is reached  with a
desired precision (usually at the 6-th digit, namely at $\simeq 10^{-6}$ per cents),
we obtain the lowest energy state of the system
(see Ref. \onlinecite{NgoSurface}).
Note that we have used several thousands of different initial conditions to check
the convergence to a single GS for each set of parameters.  Choosing $\mathbf D_{i,j}$ lying perpendicular to the
spin plane (i. e. $xz$ plane)
as indicated in Eq. (\ref{Ddef}),
we determine the GS as a function of $D$. An example is shown in Fig. \ref{ffig1}
for $\theta=\pi/6$ ($D=-0.577$) with $J_{\shortparallel}=J_\bot =1$.
We see that each spin has the same angle with its four NN in the plane
(angle between NN in adjacent planes is zero). Let us show the relation between $\theta$ and $J_{\shortparallel}$:
the energy of the spin $\mathbf S_i$ is written as
\begin{equation}
E_{i}=-4J_{\shortparallel}S^2\cos\theta-2J_\bot S^2+ 4DS^2\sin\theta
\end{equation}
where $\theta=|\theta_{i,j}|$ and care has been taken on the signs of $\sin \theta_{i,j}$ and $e_{i,j}$
when counting NN, namely
two opposite NN have opposite signs.  The minimization
of $E_i$ yields
\begin{equation}
\frac{dE_{i}}{d\theta}=0\  \ \Rightarrow \  \  -\frac{D}{J_{\shortparallel}}=\tan \theta
\  \ \Rightarrow \  \  \theta=\arctan (-\frac{D}{J_{\shortparallel}})\label{gsangle}
\end{equation}
The value of $\theta$ for a given $\frac{D}{J_{\shortparallel}}$ is precisely what obtained by
the steepest descent method.

In the present model, the DM interaction is supposed in the plane, so in the GS the angle between in-plane NN is not zero.
We show in Fig. \ref{ffig1} the relative orientation of the two NN spins in the plane.

\begin{figure}[ht!]
\centering
\includegraphics[width=6cm,angle=0]{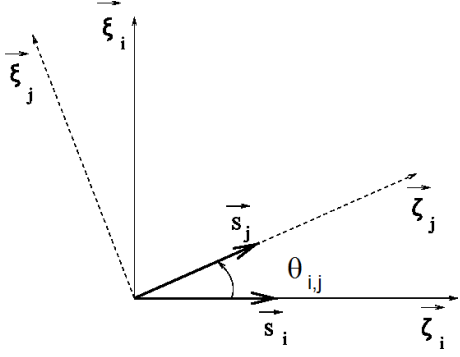}
\caption{Local coordinates in the $xz$ plane. The spin quantization axes of $\mathbf S_{i}$ and $\mathbf S_{j}$ are  $\hat\zeta_{i}$ and $\hat\zeta_{j}$ , respectively. \label{ffig2}}
\end{figure}
Note that the perpendicular axes $\hat \eta_i$ and $\hat \eta_j$ coincide. Now,
expressing the local frame of $\mathbf S_j$ in the local frame of $\mathbf S_i$, we have
\begin{eqnarray}
\hat \zeta_j&=&\cos\theta_{i,j} \hat \zeta_i + \sin \theta_{i,j}\hat \xi_i\label{local3}\\
\hat \xi_j&=&-\sin \theta_{i,j}\hat \zeta_i + \cos\theta_{i,j}\hat \xi_i\label{local4}\\
\hat \eta_j&=&\hat \eta_i\label{local5}
\end{eqnarray}
so that
\begin{eqnarray}
\mathbf S_{j}&=&S_j^x(\cos\theta_{i,j}\hat \zeta_i -\sin \theta_{i,j}\hat \xi_i)\nonumber\\
&&+S_j^y\hat \eta_i+S_j^z(\cos\theta_{i,j}\hat \zeta_i +\sin \theta_{i,j}\hat \xi_i)\label{local6}
\end{eqnarray}
The DM term of Eq. (\ref{eqn:hamil3}) can be rewritten as
\begin{eqnarray}
\mathbf S_i\wedge\mathbf S_j&=& (-S_i^zS_j^y-S_i^yS_j^x\sin \theta_{i,j}+S_i^yS_j^z\cos \theta_{i,j})\hat \xi_i\nonumber\\
&&+(S_i^xS_j^x\sin\theta_{i,j}+S_i^zS_j^z\sin\theta_{i,j})\hat \eta_i\nonumber\\
&&+(S_i^xS_j^y-S_i^yS_j^z\sin \theta_{i,j}-S_i^yS_j^x\cos \theta_{i,j})\hat \zeta_i\nonumber\\
&&\label{vpro}
\end{eqnarray}
Using Eq. (\ref{Ddef}), we have
\begin{eqnarray}
\mathcal H_{DM}&=&\sum_{\left<i,j\right>}\mathbf D_{i,j}\cdot \mathbf S_i\wedge\mathbf S_j\nonumber\\
&=&D\sum_{\left<i,j\right>}(S_i^xS_j^xe_{i,j}\sin\theta_{i,j}+S_i^zS_j^ze_{i,j}\sin\theta_{i,j})\nonumber\\
&=&\frac{D}{4}\sum_{\left<i,j\right>}[(S_i^++S_i^-)(S_j^++S_j^-)e_{i,j}\sin\theta_{i,j}\nonumber\\
&&+4S_i^zS_j^ze_{i,j}\sin\theta_{i,j}]\nonumber\\
&&\label{DMterm}
\end{eqnarray}
where we have replaced $S^x=(S^++S^-)/2$.
Note that $e_{i,j}\sin\theta_{i,j}$ is always positive since for a NN on the positive axis direction,
$e_{i,j}=1$ and
$\sin\theta_{i,j}=\sin \theta$ where $\theta$ is positively defined, while for a NN on the negative axis direction,
$e_{i,j}=-1$ and $\sin\theta_{i,j}=\sin(-\theta)=-\sin\theta$.

Note that for non collinear spin configurations, the local spin coordinates allow one to use the commutation relations between spin operators of a spin which are valid only when the $z$ spin component is defined on its quantification axis. This method has been applied for helimagnets \cite{Harada,Diep89,Quartu1998}.

\section{Self-consistent Green's function method: formulation}\label{GF}
The GF method has been developed for
non collinear
surface spin configurations in thin films \cite{NgoSurface,NgoSurface2,Diep2015,Sahbi}.
Let us briefly recall
here the principal steps of calculation and give the results for the present model.
Expressing the Hamiltonian in the local coordinates, we obtain
\begin{eqnarray}
\mathcal H &=& - \sum_{<i,j>}
J_{i,j}\Bigg\{\frac{1}{4}\left(\cos\theta_{i,j} -1\right)
\left(S^+_iS^+_j +S^-_iS^-_j\right)\nonumber\\
&+& \frac{1}{4}\left(\cos\theta_{i,j} +1\right) \left(S^+_iS^-_j
+S^-_iS^+_j\right)\nonumber\\
&+&\frac{1}{2}\sin\theta_{i,j}\left(S^+_i +S^-_i\right)S^z_j
-\frac{1}{2}\sin\theta_{i,j}S^z_i\left(S^+_j
+S^-_j\right)\nonumber\\
&+&\cos\theta_{i,j}S^z_iS^z_j\Bigg\}\nonumber\\
&+&\frac{D}{4}\sum_{\left<i,j\right>}[(S_i^++S_i^-)(S_j^++S_j^-)e_{i,j}\sin\theta_{i,j}\nonumber\\
&&+4S_i^zS_j^ze_{i,j}\sin\theta_{i,j}]\nonumber\\
&&\label{eq:HGH2}
\end{eqnarray}
As said in the previous section, the spins lie in the $xz$ planes, each on its quantization
local $z$ axis (Fig. \ref{ffig2}).

Note that unlike the sinus term of the DM Hamiltonian, Eq. (\ref{DMterm}), the sinus terms of $\mathcal H_e$,
the 3rd line of Eq. (\ref{eq:HGH2}), are zero when summed up on opposite NN (no $e_{i,j}$ to compensate).
The 3rd line disappears therefore in the following.

At this stage it is very important to note that the standard commutation relations between spin operators
$S^z$ and $S^{\pm}$ are defined with $z$ as the spin quantization axis. In non collinear spin configurations,
calculations of SW spectrum using commutation relations
without paying attention to this are wrong.

It is known that in two dimensions (2D) there is no long-range order at finite temperature ($T$) for isotropic
spin models with short-range interaction \cite{Mermin}.
Thin films have small thickness, therefore to stabilize the ordering at finite $T$ it is
useful to add an anisotropic interaction.
We use the following anisotropy between $\mathbf S_i$
and $\mathbf S_j$ which stabilizes the angle determined above between their local quantization axes $S^z_i$ and $S^z_j$:
\begin{equation}
\mathcal H_a= -\sum_{<i,j>} I_{i,j}S^z_iS^z_j\cos\theta_{i,j}
\end{equation}
where $I_{i,j}$ is supposed to be positive, small compared to $J_{\shortparallel}$, and limited to NN.
Hereafter we take $I_{i,j}=I_1$ for NN pair in the $xz$ plane,
for simplicity. As it turns out, this anisotropy helps stabilize the ordering at finite $T$ in 2D as discussed. It helps also stabilize the SW spectrum at $T=0$ in the case of thin films but it is not necessary for 2D and 3D at $T=0$.  The total Hamiltonian is finally given by
\begin{equation}\label{totalH}
\mathcal H=\mathcal H_e+\mathcal H_{DM}+\mathcal H_a
\end{equation}


We define the following two double-time GF's in the real space
\begin{eqnarray}
G_{i,j}(t,t')&=&<<S_i^+(t);S_{j}^-(t')>>\nonumber\\
&=&-i\theta (t-t')
<\left[S_i^+(t),S_{j}^-(t')\right]> \label{green59a}\\
F_{i,j}(t,t')&=&<<S_i^-(t);S_{j}^-(t')>>\nonumber\\
&=&-i\theta (t-t')
<\left[S_i^-(t),S_{j}^-(t')\right]>\label{green60}
\end{eqnarray}
The equations of motion of these functions
read
\begin{eqnarray}
i\hbar\frac{dG_{i,j}(t,t')}{dt}&=&<\left[S_i^+(t),S_{j}^-(t')\right]>\delta(t-t')\nonumber\\
&&-<<\left[\mathcal H,S_i^+\right];S_j^->>
\label{green59a}\\
i\hbar\frac{dF_{i,j}(t,t')}{dt}&=&<\left[S_i^-(t),S_{j}^-(t')\right]>\delta(t-t')\nonumber\\
&&-<<\left[\mathcal H,S_i^-\right];S_j^->>
\label{green60}
\end{eqnarray}
For the $\mathcal H_e$ and $\mathcal H_a$ parts, the above equations of motion generate terms such as
$<<S_l^zS_i^{\pm};S_j^->>$ and $<<S_l^{\pm}S_i^{\pm};S_j^->>$. These functions can be approximated
by using the Tyablikov decoupling to reduce to the above-defined $G$ and $F$ functions:
\begin{eqnarray}
&&<<S_l^zS_i^{\pm};S_j^->>\simeq <S_l^z><<S_i^{\pm};S_j^->>\label{Tya1}\\
&&<<S_l^{\pm}S_i^{\pm};S_j^->>\simeq <S_l^{\pm}> <<S_i^{\pm};S_j^->>\simeq 0\label{Tya2}
\end{eqnarray}
The last expression is due to the fact that transverse SW motions $<S_l^{\pm}>$ are zero with time.
For the DM term, the commutation relations $[\mathcal H,S_i^{\pm}]$ give rise to the following term:
\begin{equation}
D\sum_l\sin \theta [\mp S_i^z(S_l^{+}+S_l^{-})\pm 2S_i^{\pm} S_l^z]\label{HDMC}
\end{equation}
which leads to the following type of GF's:
\begin{equation}
<<S_i^zS_l^{\pm};S_j^->>\simeq <S_i^z><<S_l^{\pm};S_j^->>\label{DMa}\\
\end{equation}
Note that we have replaced $e_{i,j}\sin\theta_{i,j}$ by $\sin \theta$ where $\theta$ is positive.
The above equation is related to $G$ and $F$ functions [see Eq. (\ref{Tya2})].
The Tyablikov decoupling scheme neglects higher-order functions.

We now introduce the following in-plane Fourier
transforms $ g_{n,n'}$ and $  f_{n,n'}$ of the $G$ and $F$ Green's functions:
\begin{eqnarray}
G_{i,j}(t,t',\omega)&=&\frac{1}{\Delta} \int_{BZ} d{\vec k_{xz}}
\mbox{e}^{-i\omega(t-t')}\nonumber\\
&&\times g_{n,n'}(\omega,\vec k_{xz})\mbox{e}^{i\vec k_{xz}.(\vec R_i-\vec R_{j})}\\
F_{i,j}(t,t',\omega)&=&\frac{1}{\Delta} \int_{BZ} d{\vec k_{xz}}
\mbox{e}^{-i\omega(
t-t')}\nonumber\\
&&\times f_{n,n'}(\omega,\vec k_{xz})\mbox{e}^{i\vec k_{xz}.(\vec R_i-
\vec R_{j})}
\end{eqnarray}
where the integral is performed in the first $xz$ Brillouin zone (BZ) of surface $\Delta$,
$\omega$ is the spin-wave frequency, $n$ and $n'$ are
the  indices of the layers along the $c$ axis to which $\vec R_i$ and $\vec R_{j}$ belong
($n=1$ being the surface layer,
$n=2$ the second layer and so on).
We finally obtain the following matrix equation
\begin{equation}
\mathbf M \left( E \right) \mathbf h = \mathbf u,
\label{eq:HGMatrix}
\end{equation}
where $\mathbf M\left(E\right)$ is a square matrix of dimension
$\left(2N \times 2N\right)$, $\mathbf h$ and $\mathbf u$ are
the column matrices which are defined as follows
\begin{equation}
\mathbf h = \left(%
\begin{array}{c}
  g_{1,n'} \\
  f_{1,n'} \\
  \vdots \\
  g_{n,n'} \\
  f_{n,n'} \\
    \vdots \\
  g_{N,n'} \\
  f_{N,n'} \\
\end{array}%
\right) , \hspace{1cm}\mathbf u =\left(%
\begin{array}{c}
  2 \left< S^z_1\right>\delta_{1,n'}\\
  0 \\
  \vdots \\
  2 \left< S^z_{N}\right>\delta_{N,n'}\\
  0 \\
\end{array}%
\right) , \label{eq:HGMatrixgu}
\end{equation}
where $E=\hbar \omega$ and $\mathbf M\left(E\right)$ is given by

\begin{widetext}
\begin{equation}
\left(%
\begin{array}{ccccccccc}
  E+A_1& B_1    & C_1& 0&0& 0& 0&0&0\\
   -B_1   & E-A_1  & 0 & -C_1 &0&0&0&0&0\\
   \cdots & \cdots & \cdots &\cdots&\cdots&\cdots&\cdots&\cdots&\cdots\\
 \cdots&0&C_{n}&0&E+A_{n}&B_n&C_{n}&0&0\\
 \cdots&0&0&-C_{n}&-B_n&E-A_{n}&0&-C_{n}&0\\
         \cdots  & \cdots & \cdots &\cdots&\cdots&\cdots&\cdots&\cdots&\cdots\\
  0& 0&0& 0& 0  & C_{N}   & 0   &E + A_{N}&B_{N}\\
  0&0&0&0& 0 & 0  & -C_{N}& -B_{N}& E-A_{N}\\
\end{array}%
\right)
\end{equation}\label{eq:HGMatrixM}
\end{widetext}
with
\begin{eqnarray}
A_{n} &=& -J_{\shortparallel}[8<S^z_n>\cos\theta (1+d_n)\nonumber\\
&&- 4 <S^z_n>\gamma (\cos\theta+1)]\nonumber\\
&&-2J_{\bot} (<S^z_{n-1}>+<S^z_{n+1}>)\nonumber\\
&&-4D\sin \theta < S^z_{n}>\gamma\nonumber\\
&&+8D\sin \theta < S^z_{n}>\label{anterm}\\
B_n &=& 4J_{\shortparallel} < S^z_{n}> \gamma (\cos\theta-1)\nonumber\\
&&-4D \sin \theta < S^z_{n}>\gamma\label{bnterm}\\
C_n &=& 2J_{\bot} < S^z_{n}> \label{cnterm}
\end{eqnarray}
where $n=1,2,...,N$, $d_n=I_1/J_{\shortparallel}$, $\gamma=(\cos k_xa+\cos k_za)/2$,
$k_{x}$ and  $k_{z}$
denote the wave-vector components in the $xz$ planes, $a$ the lattice constant.
Note that (i) if $n=1$ (surface layer) then there are no  $n-1$ terms in the matrix coefficients, (ii) if $n=N$
then there are no  $n+1$ terms.  Besides, we have distinguished the in-plane NN interaction $J_{\shortparallel}$
from the inter-plane NN one $J_\bot$.

In the case of a thin film, the SW eigenvalues at a given wave vector $\vec k=(k_x,k_z)$
are calculated by diagonalizing the matrix \label{eq:HGMatrixM}.

The layer magnetization of the layer $n$ is given by (see technical details in Ref. \onlinecite{DiepTM}):
\begin{equation}\label{lm2}
\langle S_{n}^z\rangle=\frac{1}{2}-
   \frac{1}{\Delta}
   \int
   \int dk_xdk_z
   \sum_{i=1}^{2N}\frac{Q_{2n-1}(E_i)}
   {\mbox{e}^{E_i/k_BT}-1}
\end{equation}
where $n=1,...,N$, and $Q_{2n-1}(E_i)$ is the determinant obtained by replacing the $(2n-1)$-th column of
$\mathbf M$ by $\mathbf u$ at $E_i$.

The layer magnetizations can be calculated at finite temperatures self-consistently using the above formula.
The numerical method to carry out this task has been described in details in Refs. \onlinecite{Diep2015}. One can summarize here: (i) using a set of trial values (inputs) for $\langle S_{n}^z\rangle$ ($n=1,...,N)$, one diagonalizes the matrix to find spin-wave energies $E_i$ which are used to calculate the outputs $\langle S_{n}^z\rangle$ ($n=1,...,N)$ by using Eq. (\ref{lm2}), (ii) using the outputs as inputs to iterate the equations, (iii) if the output values are the same as the inputs within a precision (usually at 0.001\%), the iteration is stopped.  The method is thus self-consistent.

The value of the spin in the layer $n$ at $T=0$ is calculated by \cite{Diep2015,DiepTM}

\begin{equation}\label{surf38}
\langle S_{n}^z\rangle(T=0)=\frac{1}{2}+
   \frac{1}{\Delta} \int \int dk_xdk_z
   \sum_{i=1}^{N}Q_{2n-1}(E_i)
\end{equation}
where the sum is performed over $N$ negative values of  $E_i$ (for positive values, the
Bose-Einstein factor in Eq. (\ref{lm2}) is equal to 0 at $T=0$).

The transition temperature $T_c$ can be calculated by letting $\langle S_{n}^z\rangle$ on the left-hand side of Eq. (\ref{lm2}) to go to zero. The energy $E_i$ tends then to zero, so that we can make an expansion of the exponential at $T=T_c$.  We have
\begin{equation}\label{TCG}
\left[\frac{1}{k_BT_c}\right]=\frac{2}{\Delta}
   \int
   \int dk_xdk_z
   \sum_{i=1}^{2N}\frac{Q_{2n-1}(E_i)}{E_i}
\end{equation}

\section{Two and three dimensions: spin-wave spectrum and magnetization}\label{2D3D}
Consider just one single $xz$ plane. The above matrix is reduced to two coupled equations
\begin{eqnarray}
(E+A_n)g_{n,n'}+B_nf_{n,n'}&=&2< S^z_{n}>\delta(n,n')\nonumber\\
-B_ng_{n,n'}+(E-A_n)f_{n,n'}&=&0
\end{eqnarray}
where $A_n$ is given by (\ref{anterm}) but without $J_\bot$ term for the 2D case considered here.
Coefficients $B_n$ and $C_n$
are given by (\ref{bnterm}) and (\ref{cnterm}) with $C_n=0$.
The poles of the GF are the eigenvalues of the SW spectrum which are given by the secular equation
\begin{eqnarray}
&&(E+A_n)(E-A_n)+B_n^2=0\nonumber\\
&&[E+A_n][E-A_n]+B_n^2=0\nonumber\\
&&E^2-A_n^2+B_n^2=0\nonumber\\
&&E=\pm \sqrt{(A_n+B_n)(A_n-B_n)}\label{SWE}
\end{eqnarray}
where $\pm$ indicate the left and right SW precessions.
Several remarks are in order:

(i) if $\theta=0$, we have  $B_n=0$ and the last three terms of $A_n$ are zero. We recover then the ferromagnetic SW dispersion relation

\begin{equation}
E=2ZJ_{\shortparallel}<S_n^z>(1-\gamma)
\end{equation}
where $Z=4$ is the coordination number of the
square lattice (taking $d_n=0$),

(ii)  if $\theta=\pi$, we have $A_n=8J_{\shortparallel}<S_n^z>$, $B_n=-8J_{\shortparallel}<S_n^z>\gamma$.
We recover then the antiferromagnetic SW dispersion relation
\begin{equation}
E=2ZJ_{\shortparallel}<S_n^z>\sqrt{1-\gamma^2}
\end{equation}

(iii) in the presence of a DM interaction, we have $0<\cos \theta < 1$ ($0<\theta<\pi/2$). If $d_n=0$, the
quantity in the square root of Eq. (\ref{SWE}) is always $\geq 0$ for any $\theta$. It is zero at $\gamma=1$.
The SW spectrum is therefore stable at the long-wavelength limit. The anisotropy $d_n$ gives a gap at $\gamma=1$.


As said earlier, the necessity to include an anisotropy has a double purpose: it permits a gap and stabilizes a long-range ordering at finite $T$ in 2D systems.

\begin{figure}[h!]
\center
\includegraphics[width=6cm]{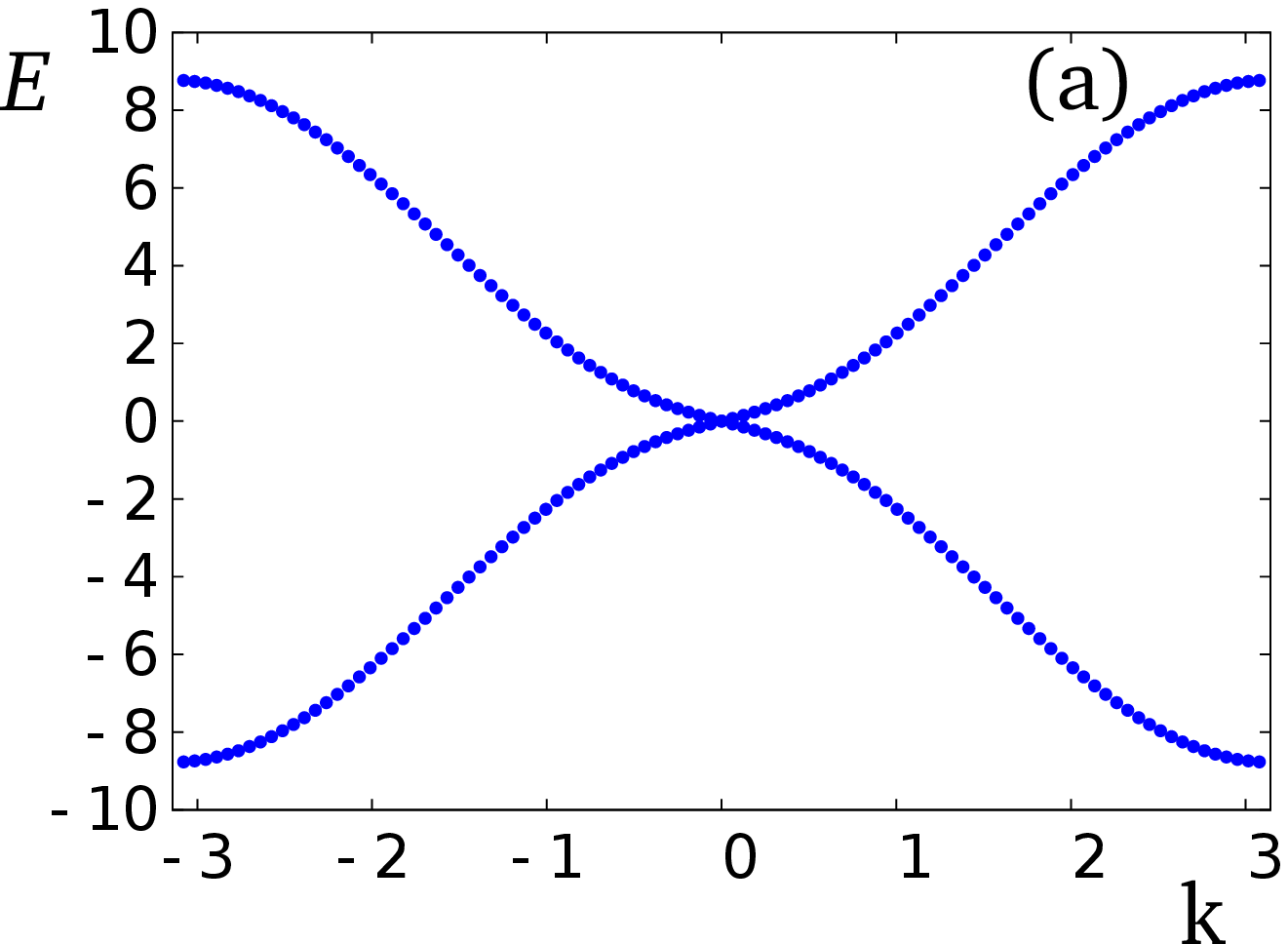}
\includegraphics[width=6cm]{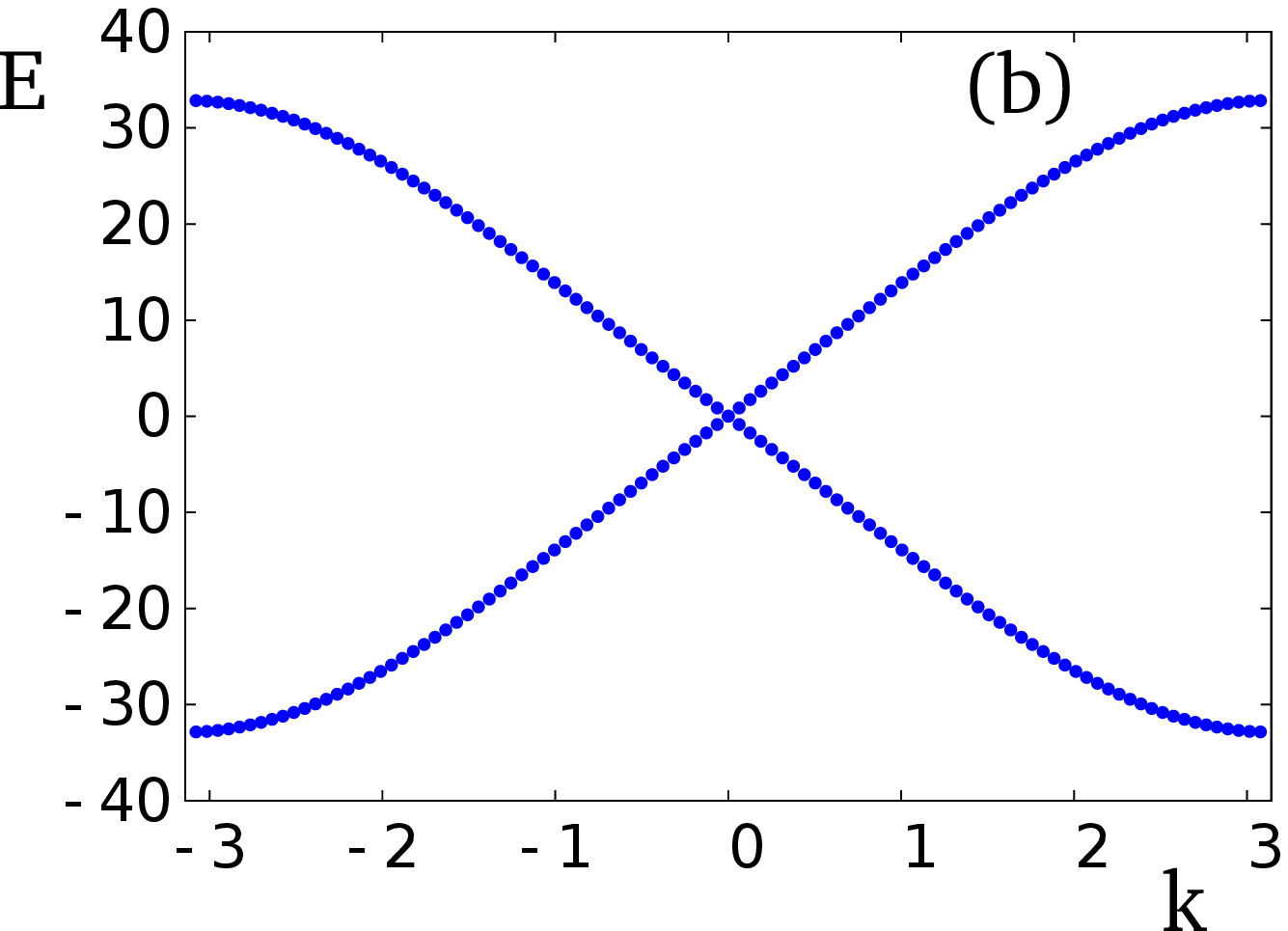}
\caption{Spin-wave spectrum $E (k)$ versus $k\equiv k_x=k_z$ for (a) $\theta=0.524$ radian and (b) $\theta=1.393$ in two dimensions at $T=0.1$.
Positive and negative branches correspond to
right and left precessions. A small $d$ ($= 0.001$) has been used to stabilized the ordering at finite $T$ in 2D. See text for comments.
\label{ffig3}}
\end{figure}

Figure \ref{ffig3} shows the SW spectrum calculated from Eq. (\ref{SWE}) for $\theta=30$ degrees ($\pi/6$ radian) and $80$ degrees (1.396 radian). The spectrum is symmetric for positive and negative wave vectors and for left
and right precessions. Note that for small $\theta$ (i. e. small $D$) $E$ is proportional to $k^2$ at low $k$ (cf. Fig. \ref{ffig3}a), as in ferromagnets. However, as $\theta$ increases, we observe that $E$ becomes linear in $k$ as seen in Fig. \ref{ffig3}b. This is similar to antiferromagnets.  The change of behavior is progressive with increasing $\theta$, we do not observe a sudden transition from $k^2$ to $k$ behavior.  This feature is also observed in three dimensions (3D) and in thin films as seen below.


\begin{figure}[ht!]
\centering
\includegraphics[width=7cm,angle=0]{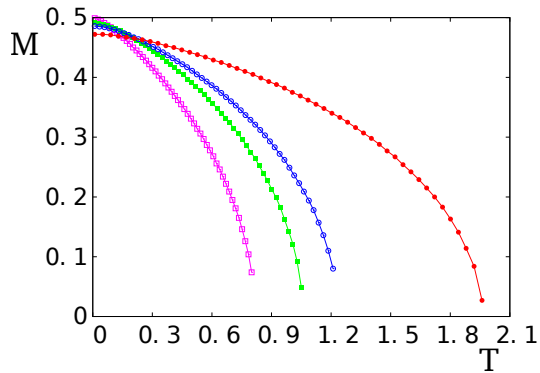}
\caption{Magnetizations $M$ versus temperature $T$ for a monolayer (2D) $\theta=0.175$ (radian), $\theta=0.524$, $\theta=0.698$,  $\theta=1.047$ (void magenta squares,  green filled squares, blue void circles and filled red circles, respectively). A small $d$ ($=0.001$) has been used to stabilized the ordering at finite $T$ in 2D. See text for comments.
\label{ffig4}}
\end{figure}

It is noted that, thanks to the existence of the anisotropy $d$, we avoid the logarithmic divergence at $k=0$ so that we can observe a long-range ordering at finite $T$
in 2D. We show in Fig. \ref{ffig4} the magnetization $M$ ($\equiv <S^z>$) calculated by Eq. (\ref{lm2}) for one layer using $d=0.001$. It is interesting to observe that $M$ depends strongly on $\theta$: at high $T$, larger $\theta$ yields stronger $M$. However, at $T=0$ the spin length is smaller for larger $\theta$ due to the so-called spin contraction \cite{DiepTM} calculated by Eq. (\ref{surf38}).  As a consequence there is a cross-over of  magnetizations with different $\theta$ at low $T$ as shown in Fig. \ref{ffig4}.


Let us study the 3D case. The crystal is periodic in three directions. We can use the Fourier transformation in the $y$ direction, namely $g_{n\pm 1}=g_n\mbox{e}^{\pm ik_ya}$ and $f_{n\pm 1}=f_n\mbox{e}^{\pm ik_ya}$.  The matrix  (\ref{eq:HGMatrixgu}) is reduced to two coupled equations of $g$ and $f$ functions, omitting index $n$,
\begin{eqnarray}
(E+A')g + Bf&=&2<S^z>\nonumber\\
-Bg+(E-A')f&=&0
\end{eqnarray}
where
\begin{eqnarray}
A' &=& -J_{\shortparallel}[8<S^z>\cos\theta (1+d)\nonumber\\
&&- 4 <S^z>\gamma (\cos\theta+1)]\nonumber\\
&&+4J_{\bot} < S^z>\cos(k_ya)\nonumber\\
&&-4D\sin \theta < S^z>\gamma\nonumber\\
&&+8D\sin \theta < S^z>\label{anterm3D}\\
B &=& 4J_{\shortparallel} < S^z> \gamma (\cos\theta-1)\nonumber\\
&&-4D \sin \theta < S^z>\gamma\label{bnterm3D}
\end{eqnarray}
The spectrum is given by
\begin{equation}\label{3Dspect}
E=\pm\sqrt{(A'+B)(A'-B)}
\end{equation}
If $\cos \theta=1$ (ferromagnetic), one has $B=0$. By regrouping the Fourier transforms in three directions,   one obtains the 3D ferromagnetic dispersion relation $E=2Z<S^z>(1-\gamma^2)$ where $\gamma=[\cos (k_xa)+\cos (k_ya)+\cos (k_za)]/3$
and $Z=6$, coordination number of the simple cubic lattice.  Unlike the 2D case where the angle is inside the plane so that the antiferromagnetic case can be recovered by setting $\cos \theta=-1$ as seen above, one cannot use the above formula to find the antiferromagnetic case because in the 3D formulation it was supposed a ferromagnetic coupling between planes, namely there is no angle between adjacent planes in the above formulation.

\begin{figure}[ht!]
\centering
\includegraphics[width=6cm,angle=0]{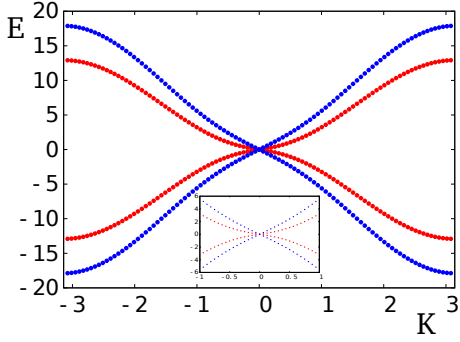}
\caption{Spin-wave spectrum $E (k)$ versus $k\equiv k_x=k_z$ for  $\theta=\pi/6$ (red circles) and  $\theta=\pi/3$ (blue circles) in three dimensions at $T=0.1$, with $d=0$. Note the linear-$k$ behavior at low $k$ for the large value of $\theta$ (inset). See text for comments.
\label{ffig5}}
\end{figure}

The same consideration as in the 2D case treated above shows that for $d=0$ the spectrum $E\geq 0$ for positive precession and $E\leq 0$ for negative precession, for any $\theta$. The limit $E=0$ is at $\gamma=1$ ($\vec k=0$). Thus there is no instability  due to the DM interaction.
 Using Eq. (\ref{3Dspect}), we have calculated the 3D spectrum. This is shown in Fig. \ref{ffig5} for a small and a large value of $\theta$. As in the 2D case, we observe $E\propto k$ when $k\rightarrow 0$ for large $\theta$.
Main properties of the system are dominated by the in-plane DM behavior.

\begin{figure}[ht!]
\centering
\includegraphics[width=6cm,angle=0]{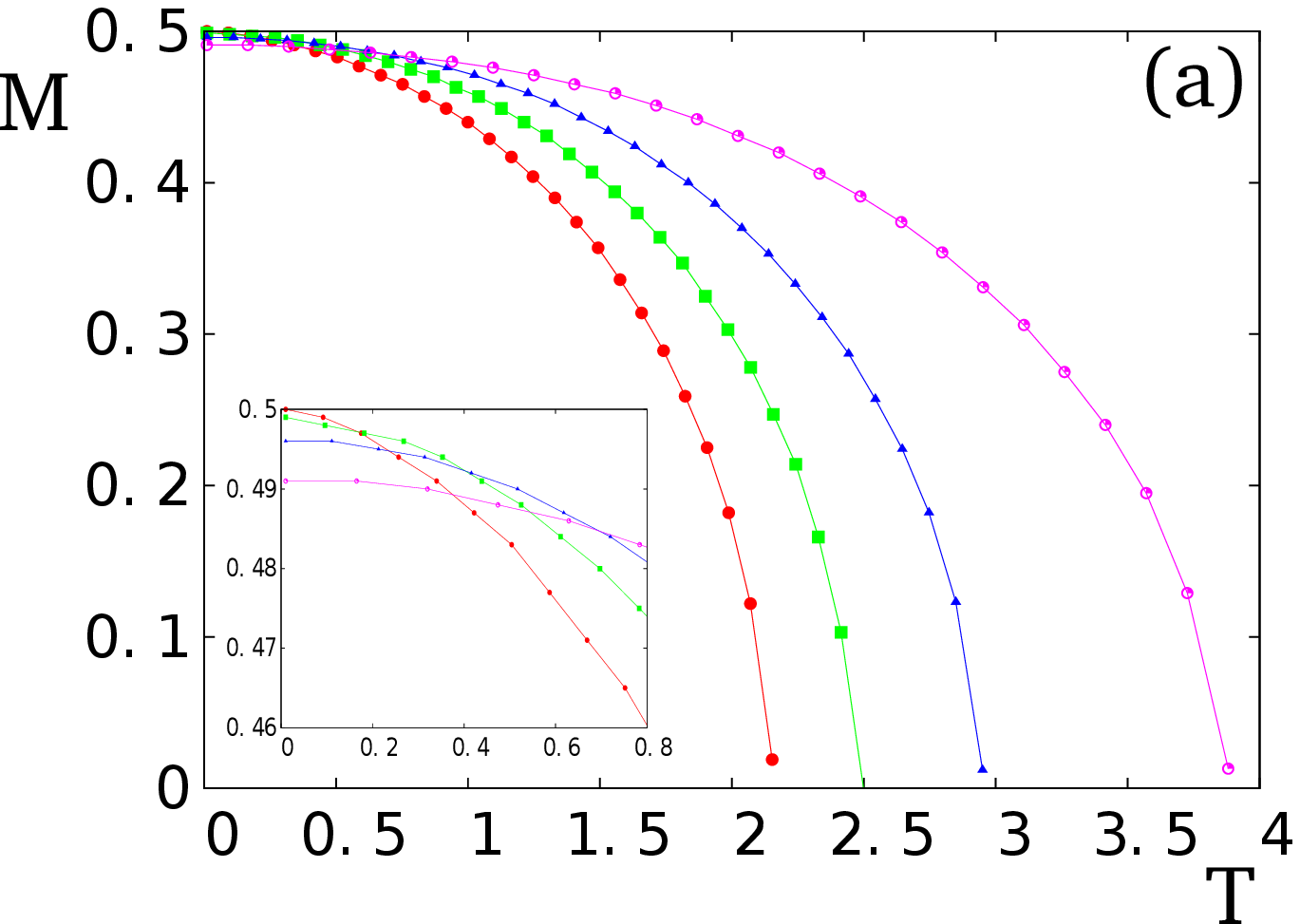}
\includegraphics[width=6.5cm,angle=0]{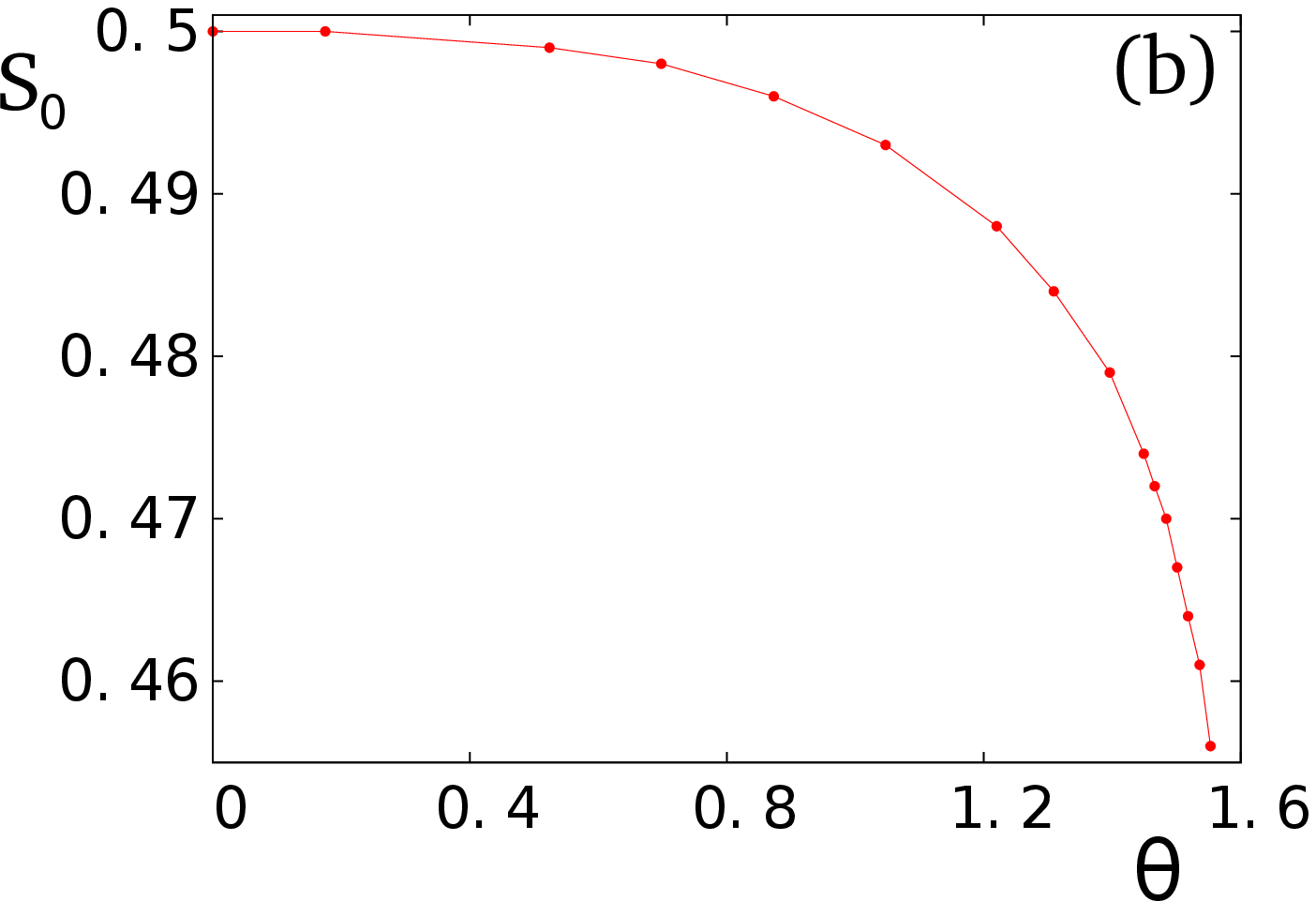}
\caption{(a) Magnetization $M$ versus temperature $T$ for a 3D crystal $\theta=0.175$ (radian), $\theta=0.524$,  $\theta=0.785$, $\theta=1.047$ (red circles, green squares, blue triangles and void magenta circles,  respectively), with $d=0$. Inset: Zoom showing the cross-over of magnetizations at low $T$ for different $\theta$, (b) The spin length $S_0$ at $T=0$ versus $\theta$. See text for comments.
\label{ffig6}}
\end{figure}

Figure \ref{ffig6}a displays the magnetization $M$ versus $T$ for several values of $\theta$.  As in the 2D case, when $\theta$ is not zero, the spins have a contraction at $T=0$: a stronger $\theta$ yields a stronger contraction. This generates a magnetization cross-over at low $T$ shown in the inset of Fig. \ref{ffig6}a. The spin length at $T=0$ versus $\theta $ is displayed in Fig. \ref{ffig6}b. Note that the spin contraction in 3D is smaller than that in 2D. This is expected since quantum fluctuations are stronger at lower dimensions.

\section{The case of a thin film: spin-wave spectrum, layer magnetizations}\label{results}

\begin{figure}[ht!]
\centering
\includegraphics[width=8cm,angle=0]{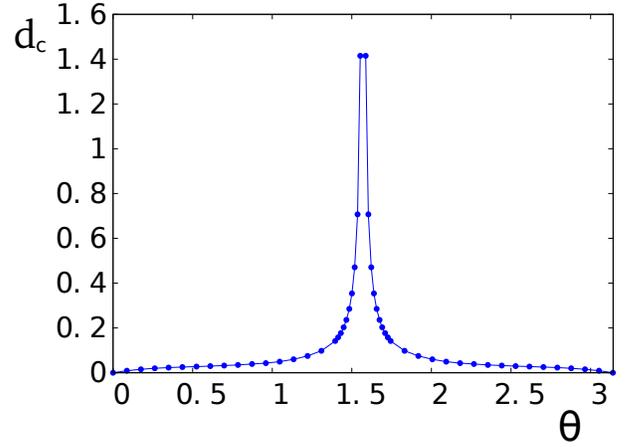}
\caption{Value $d_c$ above which the SW energy $E(\vec k=0)$ is real as a function of $\theta$ (in radian), for a 4-layer film.  Note that no spin-wave excitations are possible near the perpendicular configuration $\theta=\pi/2$.
See text for
comments. \label{ffig7}}
\end{figure}

In the 2D and 3D cases shown above, there is no need at $T=0$ to use a small anisotropy $d$. However in the case of thin films shown below, due to the lack of neighbors at the surface, the introduction of a DM interaction destabilizes the spectrum at long wave-length $\vec k=0$.  Depending on $\theta$,
we have to use a value for $d_n$
larger or equal to a "critical value" $d_c$ to avoid imaginary SW energies at $\vec k=0$.
The critical value $d_c$ is shown in Fig. \ref{ffig7} for a 4-layer film.  Note that at the perpendicular configuration $\theta=\pi/2$, no SW excitation is possible: SW cannot propagate in a perpendicular spin configuration since the wave-vectors cannot be defined.

\begin{figure}[ht!]
\centering
\includegraphics[width=6cm,angle=0]{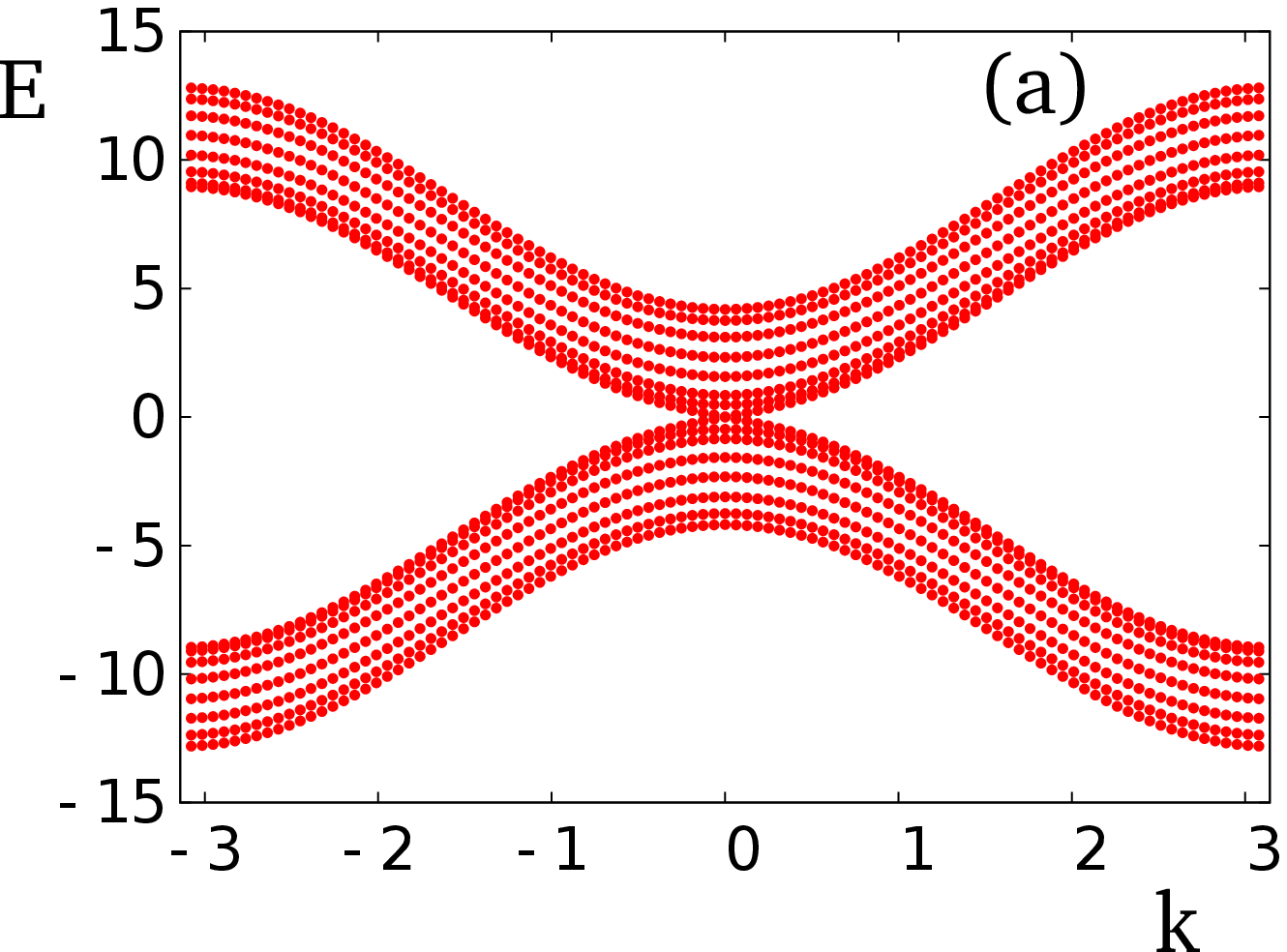}
\includegraphics[width=6cm,angle=0]{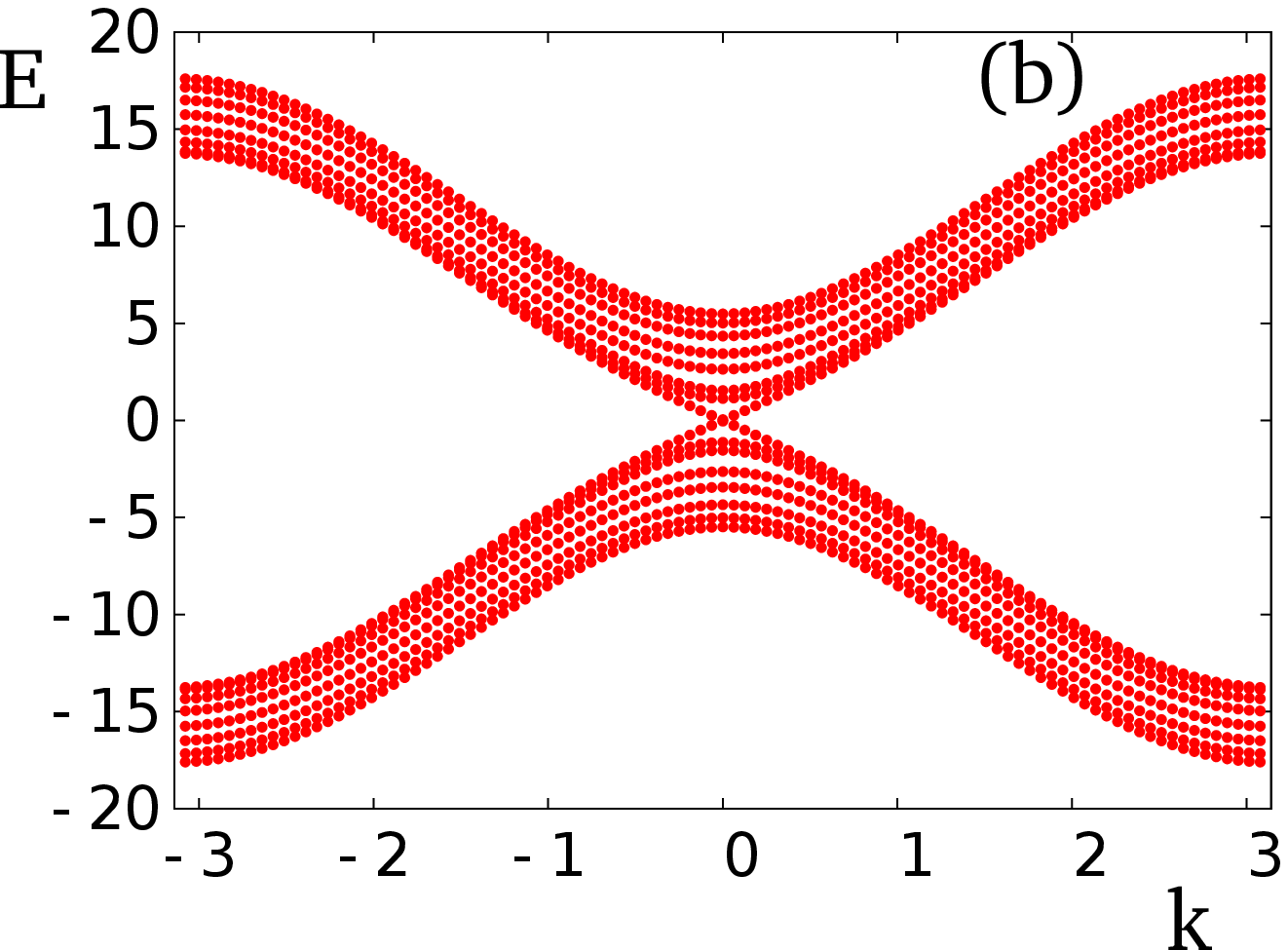}
\caption{Spin-wave spectrum $E (k)$ versus $k\equiv k_x=k_z$ for a thin
film of 8 layers: (a) $\theta=\pi/6$ (in radian) (b) $\theta=\pi/3$, using $d=d_c$ for each case ($d_c=0.012$ and 0.021, respectively).
Positive and negative branches correspond to right and left precessions.
Note the linear-$k$ behavior at low $k$ for the large $\theta$ case. See text for comments.
\label{ffig8}}
\end{figure}

We show now a SW spectrum at a given thickness $N$.
There are 2$N$ energy values half
of them are positive and the other half negative (left and right precessions): $E_i$ ($i=1,...,2N)$.
Figure \ref{ffig8} shows the case of a film of 8 layers with $J_{\shortparallel}=J_\bot=1$
for a weak and a strong value of $D$ (small and large $\theta$). As in the 2D and 2D cases, for strong $D$, $E$ is proportional to $k$ at small $k$
(cf. Fig. \ref{ffig8}b). It is noted that this behavior concerns only the first mode. The upper modes remain in the $k^2$ behavior.

\begin{figure}[ht!]
\centering
\includegraphics[width=6cm,angle=0]{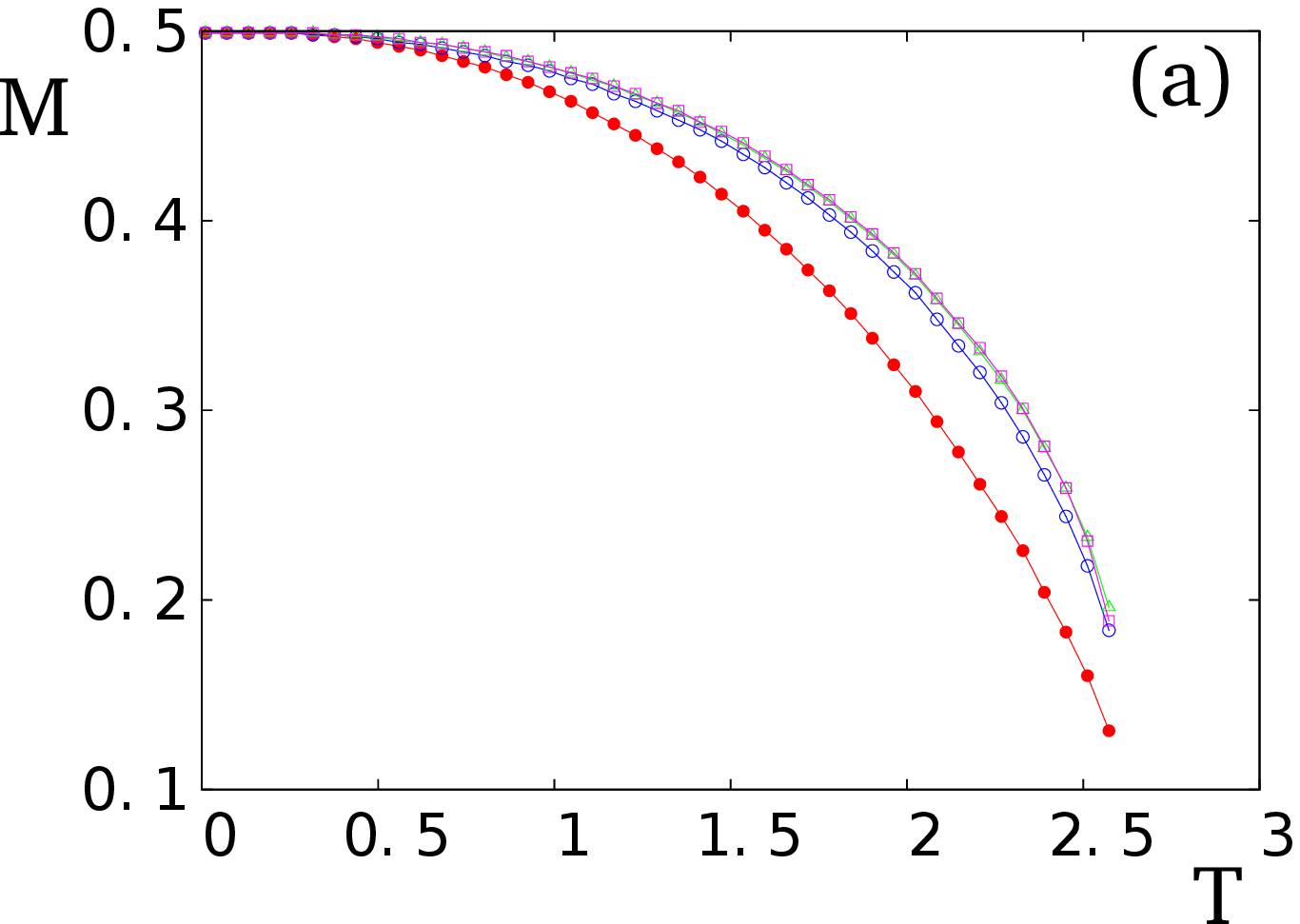}
\includegraphics[width=6cm,angle=0]{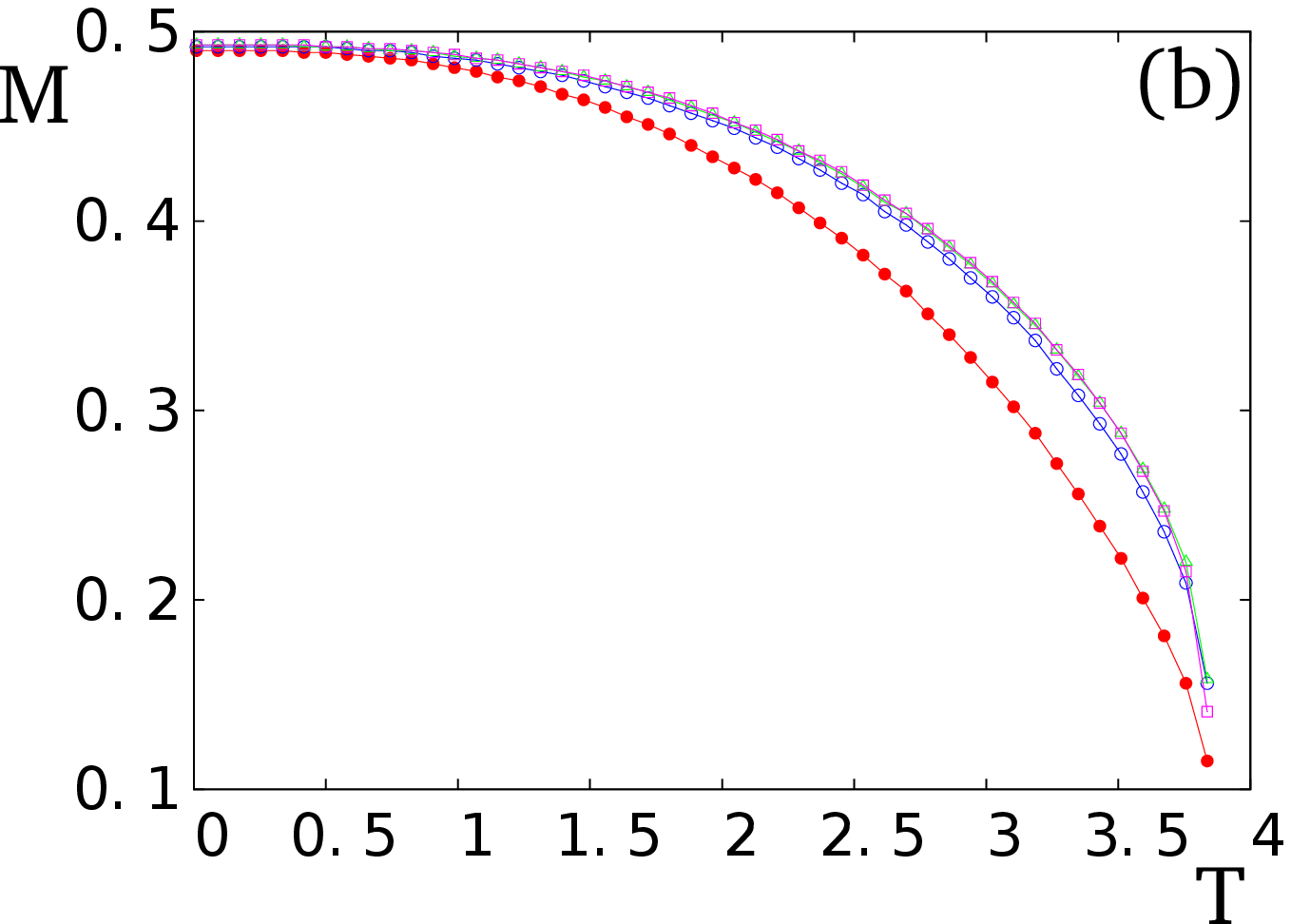}
\caption{8-layer film: layer magnetizations $M$ versus temperature $T$ for (a) $\theta=\pi/6$ (radian), (b) $\theta=\pi/3$, with $d=0.1$.
 Red circles, blue void circles, green void triangles and magenta squares correspond respectively to the first, second, third and fourth layer.
\label{ffig9}}
\end{figure}

Figure \ref{ffig9} shows the layer magnetizations of the first four layers
in a 8-layer film (the other half is symmetric) for several values of $\theta$.
In each case, we see that the surface layer magnetization is smallest.
This is a general effect of the lack of neighbors for surface spins even when there is no surface-localized SW as in the present simple-cubic lattice case \cite{DiepTM}.

\begin{figure}[h!]
\center
\includegraphics[width=6cm]{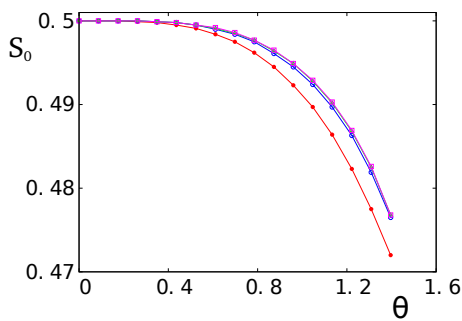}
\caption{Spin length $S_0$ at $T=0$ of the first 4 layers as a function of $\theta$, for $N=8$, $d=0.1$.   Red circles, blue void circles, green void triangles and magenta squares correspond respectively to the first, second, third and fourth layer.\label{ffig10}}
\end{figure}

The spin length at $T=0$ for a 8-layer film is shown in Fig. \ref{ffig10} as a function of $\theta$.
One observes that the spins are strongly contracted with large $\theta$.

\begin{figure}[h!]
\center
\includegraphics[width=6cm]{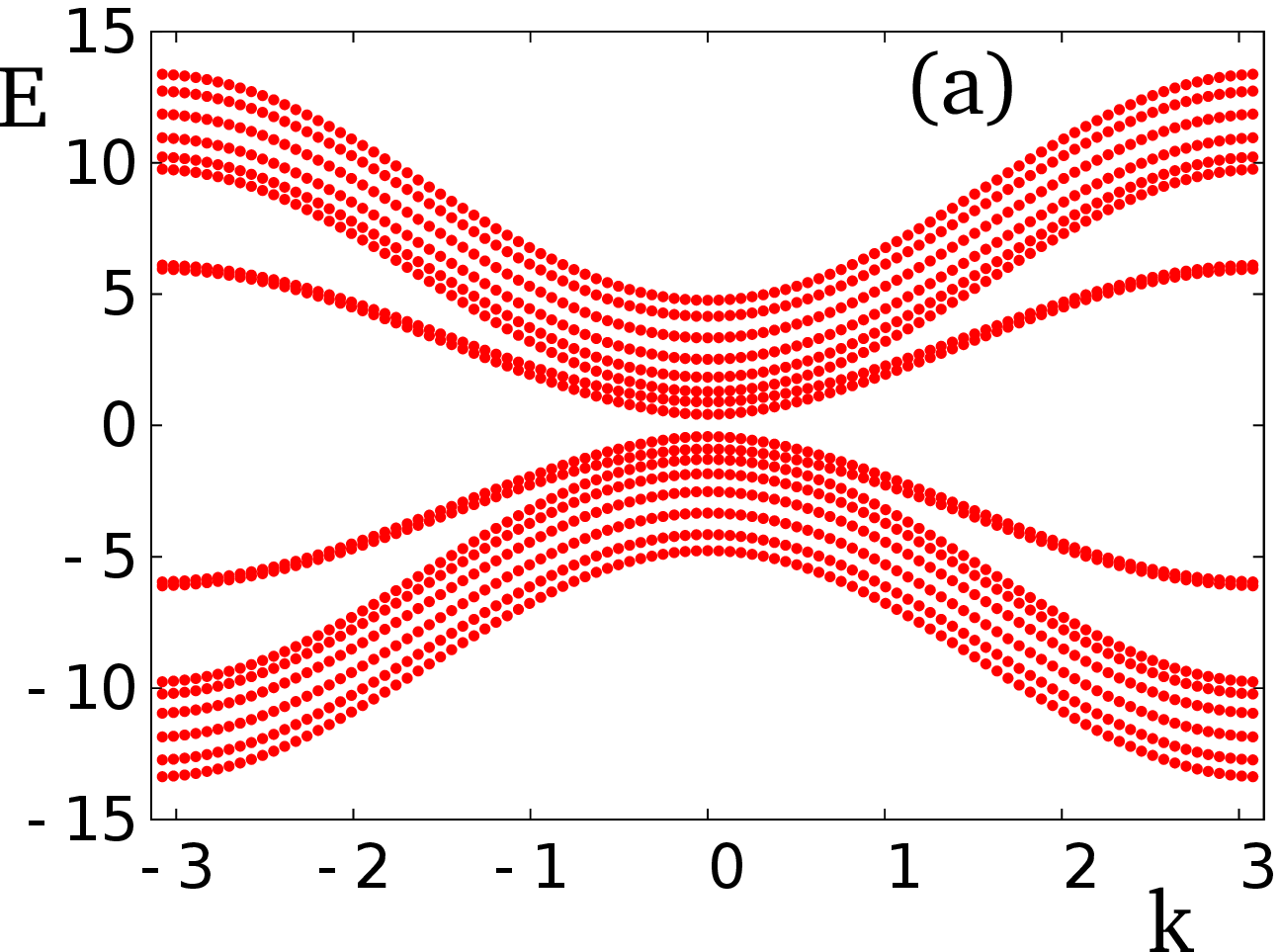}
\includegraphics[width=6.5cm]{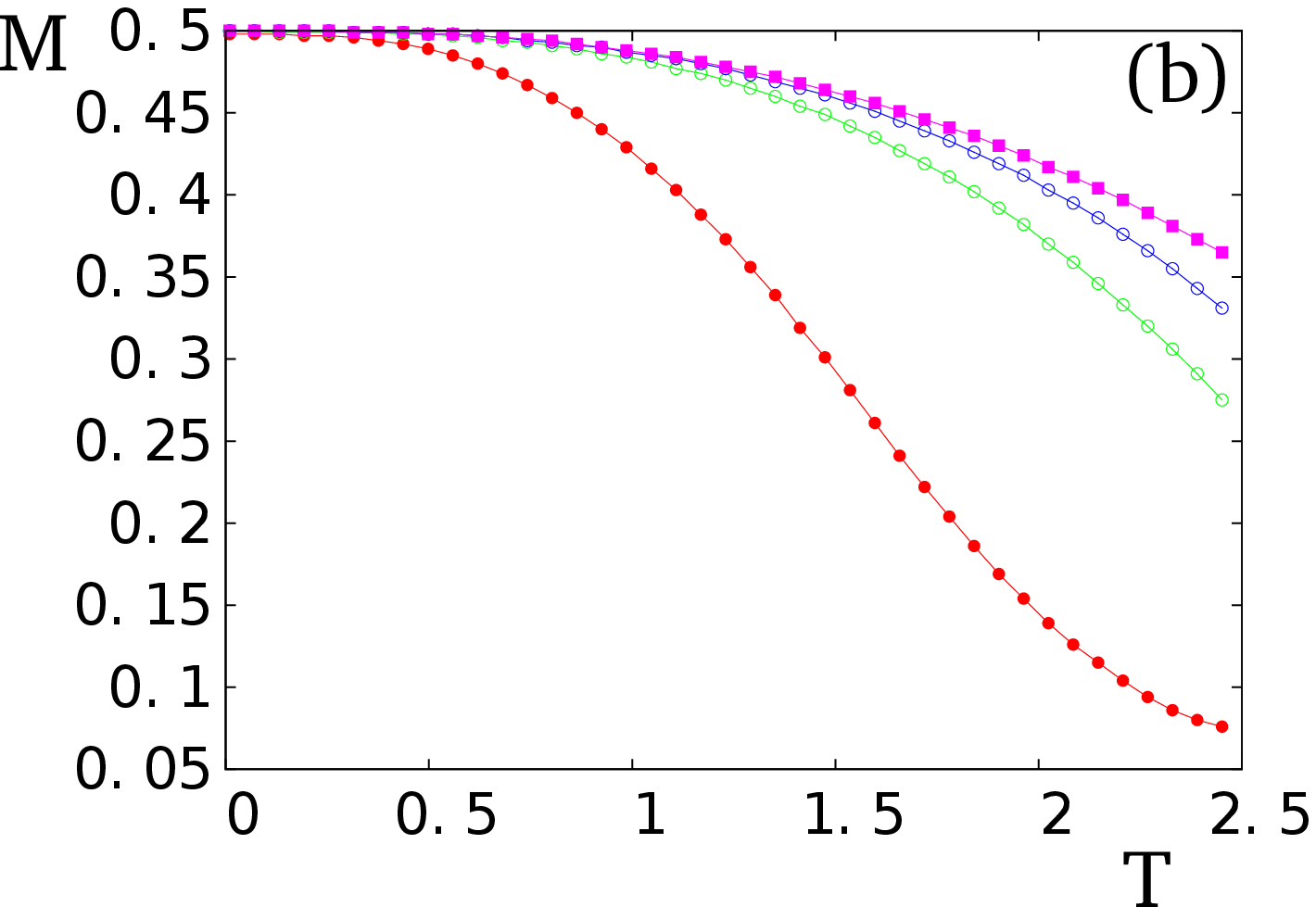}
\caption{Surface effect: (a) spin-wave spectrum $E(k)$ versus $k=k_x=k_z$ for a thin film of 8 layers: $\theta=\pi/6$,  $d=0.2$, $J^s_{\shortparallel}=0.5$, $J^s_{\bot}=0.5$, the gap at $k=0$ is due to $d$. The surface-mode branches are detached from the bulk spectrum. (b) Layer magnetizations versus $T$ for the first, second, third and  fourth layer (red circles, green void circles, blue void circles and magenta filled squares, respectively). See text for comments.\label{ffig11}}
\end{figure}

Let us touch upon the surface effect in the present model.  We know that for the simple cubic lattice, if the interactions are the same everywhere in the film, then there is no surface localized modes, and this is true with DM interaction (see spectrum in Fig. \ref{ffig8}) and without DM interaction (see Ref. \onlinecite{Diep1979}).  In order to create surface modes, we have to take the surface exchange interactions different from the bulk ones. Low-lying branches of surface modes which are "detached" from the bulk spectrum are seen in the SW spectrum shown in Fig. \ref{ffig11}a with $J^s_{\shortparallel}=0.5$, $J^s_{\bot}=0.5$. These surface modes strongly affect the surface magnetization as observed in Fig. \ref{ffig11}b: the surface magnetization is strongly diminished with increasing $T$.  The role of surface-localized modes on the strong decrease of the surface magnetization as $T$ increases has already been analyzed more than 30 years ago \cite{Diep1979}.

\begin{figure}[h!]
\center
\includegraphics[width=7cm]{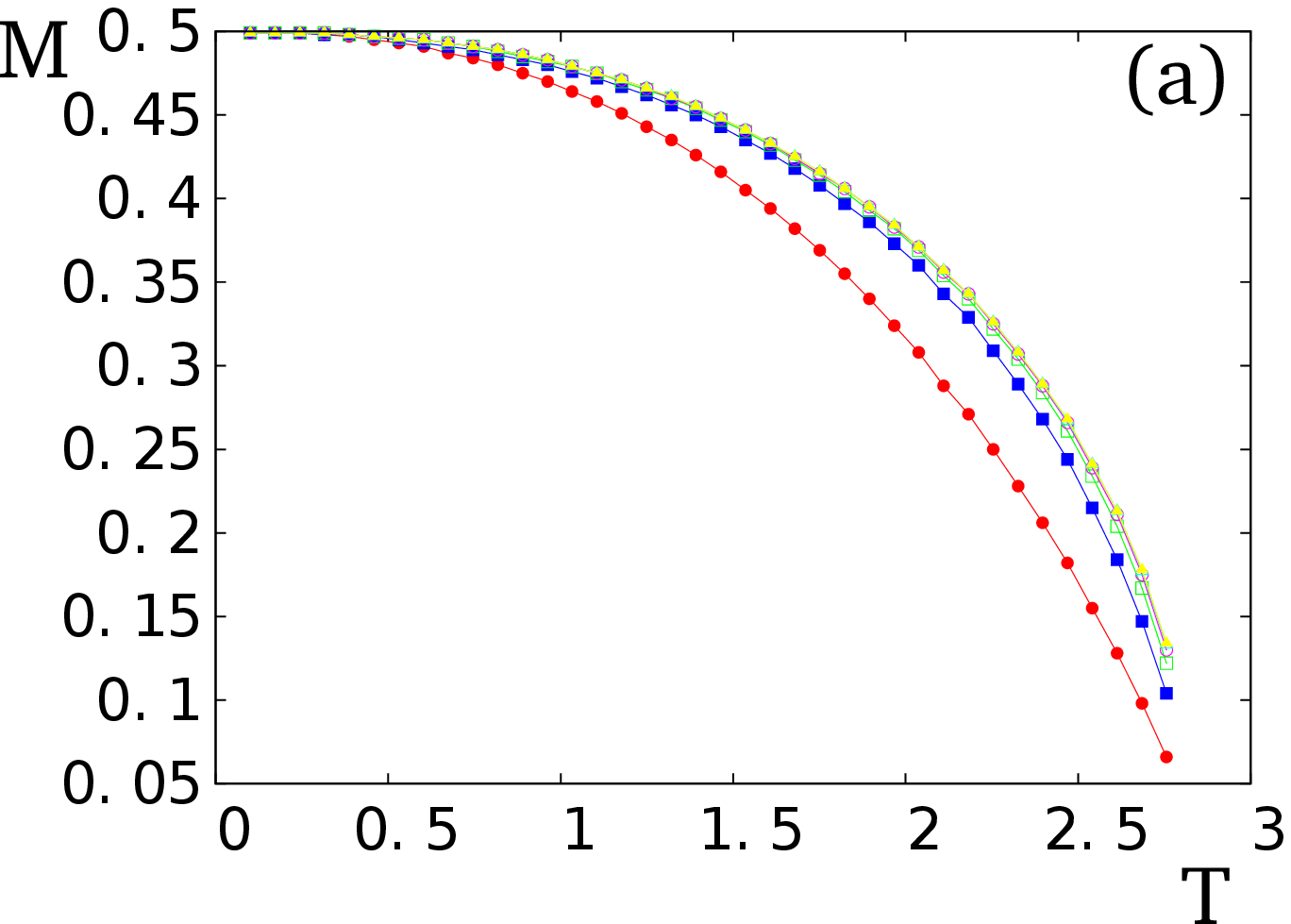}
\includegraphics[width=7cm]{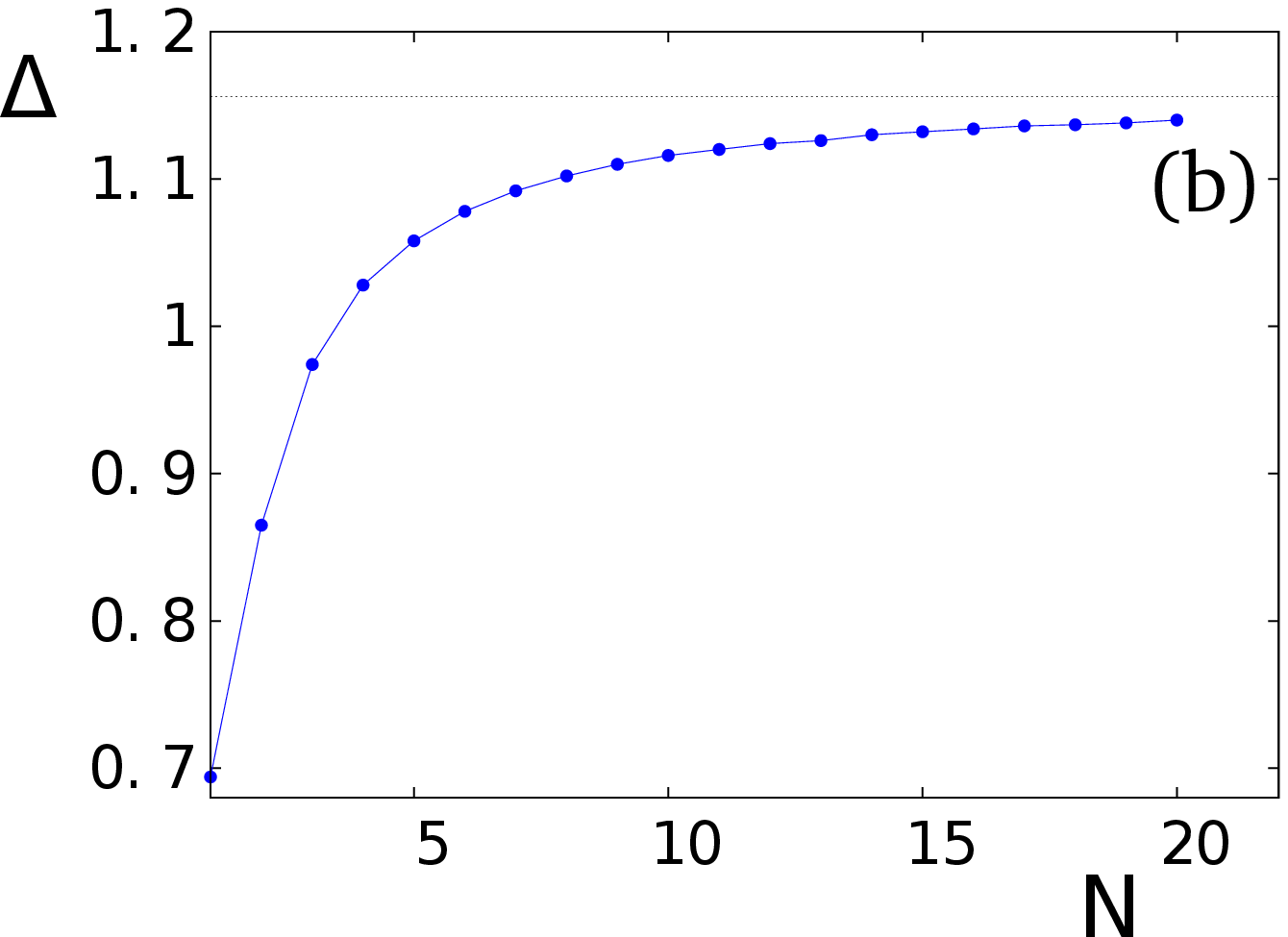}
\includegraphics[width=7cm]{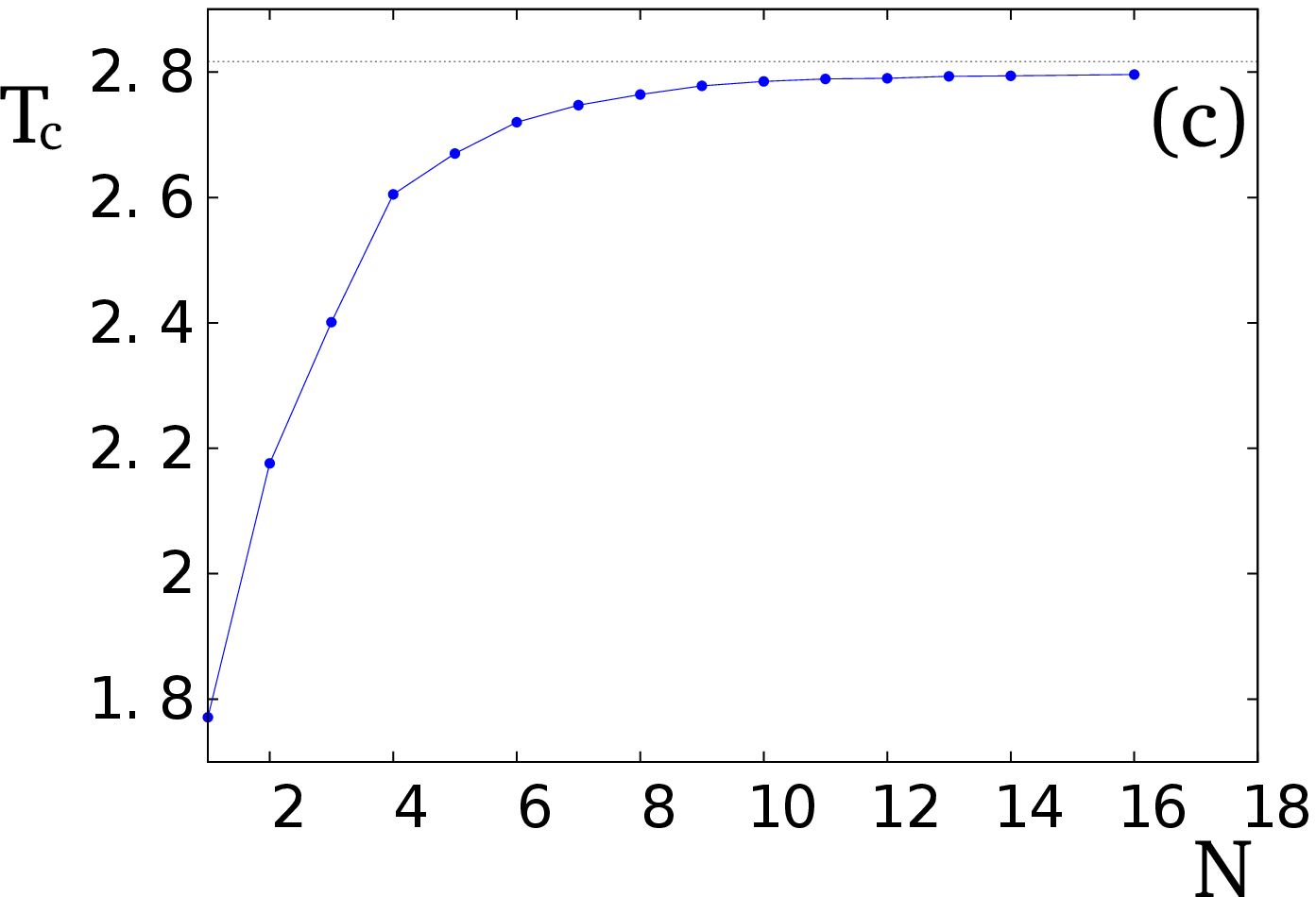}
\caption{12-layer film: (a) Layer magnetizations versus $T$ for $\theta=\pi/6$ and anisotropy $d=0.1$. Red circles, blue squares, green void squares magenta circles, void turquoise triangles and brown triangles correspond respectively to first, second, third, fourth, fifth and sixth layer, (b) Gap at $k=0$ as a function of film thickness $N$ for $\theta=\pi/6$, $d=0.1$, at $T=0.1$, (c) Critical temperature $T_c$ versus the film thickness $N$ calculated with $\theta=\pi/6$ and $d=0.1$ using
Eq. (\ref{TCG}). Note that for infinite thickness (namely 3D), $T_c\simeq 2.8$ for $d=0.1$.\label{ffig12} }
\end{figure}

We show now the effect of the film thickness in the present model. The case of thickness $N=12$ is shown in Fig. \ref{ffig12}a with $\theta=\pi/6$ where the layer magnetizations versus $T$ are shown in details. The gap at $k=0$ due to $d$ is shown in Fig. \ref{ffig12}b as a function of the film thickness $N$ for $d=0.1$ and $\theta=\pi/6$, at $T=0$. We see that the gap depends not only on $d$ but also on the value of the surface magnetization which is larger for thicker films.
The transition temperature $T_c$ versus the thickness $N$ is shown in Fig. \ref{ffig12}c where one observes that $T_c$  tends rapidly to the bulk value (3D) which is $\simeq 2.82$ for $d=0.1$.

%
%

\section{Concluding remarks}\label{concl}

By a self-consistent Green's function theory, we obtain the expression of the spin-wave dispersion relation in 2D and 3D as well as in a thin film. Due to the competition between ferromagnetic interaction $J$ and the perpendicular DM interaction $D$, the GS is non linear with an angle $\theta$ which is shown to explicitly depend on the ratio $D/J$. The spectrum is shown to depend on $\theta$ and the layer magnetization is calculated self-consistently as a function of temperature up to the critical temperature $T_c$.

We have obtained new and interesting results. In particular we have showed that (i) the spin-wave excitation in 2D and 3D crystals is stable at $T=0$  with the non collinear spin configuration induced by the DM interaction $D$ without the need of an anisotropy, (ii) in the case of thin films, we need a small anisotropy $d$ to stabilize the spin-wave excitations because of the lack of neighbors at the surface, (iii) the spin-wave energy $E$ depends on $D$, namely on $\theta$: at the long wave-length limit, $E$ is proportional to $k^2$ for small $D$ but $E$ is linear in $k$ for strong $D$, in 2D and 3D as well as in a thin film, (iv) quantum fluctuations are inhomogeneous for layer magnetizations near the surface, (v) unlike in some previous works, spin waves in systems with asymmetric DM interactions are found to be symmetric with respect to opposite  propagation directions.

\acknowledgments
This study is a part of a project financed by Narodowe
Centrum Nauki (National Science Center of Poland), Grant
No. DEC-2013/08/M/ST3/00967.
SEH acknowledges  a financial support from Agence Universitaire de la Francophonie (AUF).

{}


\begin{thebibliography}{0}%
\makeatletter
\providecommand \@ifxundefined [1]{%
 \@ifx{#1\undefined}
}%
\providecommand \@ifnum [1]{%
 \ifnum #1\expandafter \@firstoftwo
 \else \expandafter \@secondoftwo
 \fi
}%
\providecommand \@ifx [1]{%
 \ifx #1\expandafter \@firstoftwo
 \else \expandafter \@secondoftwo
 \fi
}%
\providecommand \natexlab [1]{#1}%
\providecommand \enquote  [1]{``#1''}%
\providecommand \bibnamefont  [1]{#1}%
\providecommand \bibfnamefont [1]{#1}%
\providecommand \citenamefont [1]{#1}%
\providecommand \href@noop [0]{\@secondoftwo}%
\providecommand \href [0]{\begingroup \@sanitize@url \@href}%
\providecommand \@href[1]{\@@startlink{#1}\@@href}%
\providecommand \@@href[1]{\endgroup#1\@@endlink}%
\providecommand \@sanitize@url [0]{\catcode `\\12\catcode `\$12\catcode
  `\&12\catcode `\#12\catcode `\^12\catcode `\_12\catcode `\%12\relax}%
\providecommand \@@startlink[1]{}%
\providecommand \@@endlink[0]{}%
\providecommand \url  [0]{\begingroup\@sanitize@url \@url }%
\providecommand \@url [1]{\endgroup\@href {#1}{\urlprefix }}%
\providecommand \urlprefix  [0]{URL }%
\providecommand \Eprint [0]{\href }%
\providecommand \doibase [0]{http://dx.doi.org/}%
\providecommand \selectlanguage [0]{\@gobble}%
\providecommand \bibinfo  [0]{\@secondoftwo}%
\providecommand \bibfield  [0]{\@secondoftwo}%
\providecommand \translation [1]{[#1]}%
\providecommand \BibitemOpen [0]{}%
\providecommand \bibitemStop [0]{}%
\providecommand \bibitemNoStop [0]{.\EOS\space}%
\providecommand \EOS [0]{\spacefactor3000\relax}%
\providecommand \BibitemShut  [1]{\csname bibitem#1\endcsname}%
\let\auto@bib@innerbib\@empty
\end{thebibliography}%


\begin{thebibliography}{9}

\bibitem{Dzyaloshinskii} I. E. Dzyaloshinskii, Sov. Phys. JETP {\bf 5}, 1259 (1957).
\bibitem{Moriya} T. Moriya, Phys. Rev. {\bf 120}, 91 (1960).

\bibitem{Sergienko}I. A. Sergienko and E. Dagotto,   Role of the Dzyaloshinskii-Moriya interaction in multiferroic perovskites,	Phys. Rev. B {\bf 73}, 094434 (2006).

\bibitem{Ederer} Claude Ederer. and Nicola A. Spaldin,   Weak ferromagnetism and magnetoelectric coupling in bismuth ferrite, 	Phys. Rev. B {\bf 71}, 060401(R) (2005).


\bibitem{Maleyev} Maleyev, Phys. Rev. B {\bf 73}, 174402 (2006).

\bibitem{Lin} Shi-Zeng Lin, Avadh Saxena, and Cristian D. Batista, Phys. Rev. B {\bf 91}, 224407 (2015).
\bibitem{Bogdanov} A. N. Bogdanov and D. A. Yablonskii, Sov. Phys. JETP
{\bf 68}, 101 (1989).
\bibitem{Rossler} U.K.R\"{o}ßler, A. N. Bogdanov, and C. Pfleiderer,
Nature (London) {\bf 442}, 797 (2006).
\bibitem{Muhlbauer} S. M\"{u}hlbauer, B. Binz, F. Jonietz, C. Pfleiderer, A. Rosch, A. Neubauer, R. Georgii, and P. B\"{o}ni, Science {\bf 323}, 915 (2009).
\bibitem{Yu1} X. Z. Yu, Y. Onose, N. Kanazawa, J. H. Park, J. H. Han, Y.
Matsui, N. Nagaosa, and Y. Tokura, Nature (London) {\bf 465}, 901 (2010).
\bibitem{Yu2} X. Z. Yu, N. Kanazawa, Y. Onose, K. Kimoto, W. Z. Zhang,
S. Ishiwata, Y. Matsui, and Y. Tokura, Nat. Mater. {\bf 10}, 106 (2011).
\bibitem{Seki} S. Seki, X. Z. Yu, S. Ishiwata, and Y. Tokura,
Science {\bf 336}, 198 (2012).
\bibitem{Adams} T. Adams, A. Chacon, M. Wagner, A. Bauer, G. Brandl, B.
Pedersen, H. Berger, P. Lemmens, and C. Pfleiderer,
Phys. Rev. Lett. {\bf 108}, 237204 (2012).

\bibitem{Heurich} J. Heurich, J. K\"{o}nig, and A. H. MacDonald, Phys. Rev. B {\bf 68}, 064406 (2003).

\bibitem{Wessely} O. Wessely, B. Skubic, and L. Nordstrom, Phys. Rev. B {\bf 79}, 104433 (2009).

\bibitem{Jonietz} F. Jonietz, S. M\"{u}hlbauer, C. Pfleiderer, A. Neubauer, W. Munzer, A. Bauer, T. Adams, R. Georgii, P. B\"{o}ni, R. A. Duine, K. Everschor, M. Garst, and A. Risch, Science {\bf 330}, 1648 (2010).

\bibitem{Heide} M. Heide, G. Bihlmayer, and S. Blügel, Dzyaloshinskii-Moriya interaction accounting for the orientation of magnetic domains
in ultrathin films: Fe/W(110), Phys. Rev. B {\bf 78}, 140403(R) (2008).



\bibitem{Rohart} S. Rohart. and A. Thiaville, Skyrmion confinement in ultrathin film nanostructures in the presence of
Dzyaloshinskii-Moriya interaction, Phys. Rev. B {\bf 88}, 184422 (2013).

\bibitem{Fert2013} A. Fert, V. Cros, and J. Sampaio,
Nat. Nanotechnol. {\bf 8}, 152 (2013).


\bibitem{Yoshimori} A. Yoshimori, J. Phys. Soc. Jpn {\bf 14}, 807 (1959).
\bibitem{Villain59} J. Villain, Phys. Chem. Solids {\bf 11}, 303 (1959).

\bibitem{Puszkarski} H. Puszkarski and P. E. Wigen, Effect  of Dzialoshinsky-Moriya Interactions on
Propagation of Spin Waves in Ferromagnets: Dynamical Canting, Phys. Rev. Lett. {\bf 35}, 1017 (1975).

\bibitem{Zakeri}Kh. Zakeri, Y. Zhang, J. Prokop, T.-H. Chuang, N. Sakr, W. X. Tang, and J. Kirschner,  Asymmetric Spin-Wave Dispersion on Fe(110):
Direct Evidence of the Dzyaloshinskii-Moriya Interaction,
Phys. Rev. Lett. {\bf 104}, 137203 (2010)


\bibitem{Wang} Weiwei Wang, Maximilian Albert, Marijan Beg, Marc-Antonio Bisotti, Dmitri Chernyshenko,
David Cortés-Ortuño, Ian Hawke, and Hans Fangohr, Magnon-Driven Domain-Wall Motion with the Dzyaloshinskii-Moriya Interaction, Phys. Rev. Lett. {\bf 114}, 087203 (2015).


\bibitem{Stashkevich} A. A. Stashkevich, M. Belmeguenai, Y. Roussign\'e, S. M. Cherif, M. Kostylev, M. Gabor,
D. Lacour, C. Tiusan, and M. Hehn, Experimental study of spin-wave dispersion in Py/Pt film structures in the presence
of an interface Dzyaloshinskii-Moriya interaction, Phys. Rev. B {\bf 91}, 214409 (2015).

%

\bibitem{Moon} Jung-Hwan Moon, Soo-Man Seo, Kyung-Jin Lee, Kyoung-Whan Kim, Jisu Ryu, Hyun-Woo Lee, R. D. McMichael, and M. D. Stiles, Spin-wave propagation in the presence of interfacial Dzyaloshinskii-Moriya interaction, Phys. Rev. B {\bf 88}, 184404 (2013).

\bibitem{NgoSurface} V. Thanh Ngo and H. T. Diep, Effects of frustrated surface in Heisenberg thin films,
Phys. Rev. B {\bf 75}, 035412 (2007), Selected for the Vir. J. Nan. Sci. Tech. {\bf 15}, 126 (2007).

\bibitem{Harada} I. Harada and K. Motizuki, J. Phys. Soc. Jpn {\bf 32}, 927 (1972).
\bibitem{Diep89} H. T. Diep, Low-temperature properties of quantum Heisenberg helimagnets, Phys. Rev. B {\bf 40}, 741 (1989).
\bibitem{Quartu1998} R. Quartu and H. T. Diep, Phase diagram of  body-centered tetragonal Helimagnets, J. Magn. Magn. Mater. {\bf 182}, 38 (1998).

%

\bibitem{NgoSurface2} V. Thanh Ngo and H. T. Diep, Frustration effects in antiferrormagnetic face-centered cubic Heisenberg films,
J. Phys: Condens. Matter. {\bf 19}, 386202 (2007).


\bibitem{Diep2015} H. T. Diep, Quantum Theory of Helimagnetic Thin Films, Phys. Rev. B {\bf 91}, 014436 (2015).


\bibitem{Sahbi} Sahbi El Hog and H. T. Diep, Helimagnetic Thin Films: Surface Reconstruction, Surface Spin-Waves, Magnetization, J. Magn. and Magn. Mater.  400, 276-281 (2016).


\bibitem{Mermin} N. D. Mermin and H. Wagner, Phys. Rev. Lett. {\bf 17}, 1133 (1966).






%
%
%
%
%
%
%

\bibitem{DiepFSS} H. T. Diep (ed.), {\it Frustrated Spin Systems}, 2nd edition, World Scientific (2013).




\bibitem{DiepTM} See for example H. T. Diep, {\it Theory of Magnetism: Application to Surface Physics}, World Scientific, Singapore (2014).

\bibitem{Diep1979} Diep-The-Hung, J. C. S. Levy and O. Nagai,
Effect of surface spin-waves and surface anisotropy in magnetic thin films at finite temperatures, Phys. Stat. Solidi (b) {\bf 93}, 351 (1979).


%
%
%
%
%
%
%

\end{thebibliography}
\end{document}